\documentclass[twocolumn,showpacs,preprintnumbers,amsmath,amssymb,prb,superscriptaddress]{revtex4}

\usepackage{graphicx}
\usepackage{color}
\graphicspath{{figures/}}

\newcommand{\bra}[1]{\left\langle #1 \right|}
\newcommand{\ket}[1]{\left| #1 \right\rangle}
\newcommand{\expct}[2]{\left\langle #1 \right\rangle_{#2}}
\newcommand{\expcts}[2]{\langle #1 \rangle_{#2}}
\newcommand{\cumu}[2]{\langle\!\langle #1 \rangle\!\rangle_{#2}}

\newcommand{\ti}[1]{ {\tilde #1} }
\newcommand{\tx}{ {\tilde x} }
\newcommand{\tv}{ {\tilde v} }
\newcommand{\tA}{ {\tilde A} }

\newcommand{\gext}{\gamma_\mathrm{ext}}
\newcommand{\tgext}{\tilde{\gamma}_\mathrm{ext}}
\newcommand{\nosc}{n_\mathrm{osc}}
\newcommand{\osc}{\mathrm{osc}}
\newcommand{\TB}{T_B}
\newcommand{\tTB}{\ti{T}_B}
\newcommand{\geff}{\gamma_\mathrm{eff}}
\newcommand{\Teff}{T_\mathrm{eff}}
\newcommand{\gSSET}{\gamma_\mathrm{SSET}}
\newcommand{\TSSET}{T_\mathrm{SSET}}

\usepackage{ulem}
\renewcommand{\emph}[1]{\textit{#1}}

 \DeclareMathOperator{\Tr}{Tr}
 \DeclareMathOperator{\hc}{h.c.}

\begin{document}

\title{Transport properties of a superconducting single-electron transistor\\ 
coupled to a nanomechanical oscillator}
\author{V.~Koerting}
\thanks{These authors contributed equally to this work.}
\affiliation{Department of Physics, University of Basel, CH-4056
  Basel, Switzerland}
\author{T.~L.~Schmidt}
\thanks{These authors contributed equally to this work.}
\affiliation{Department of Physics, University of Basel, CH-4056
  Basel, Switzerland}
\author{C.~B.~Doiron}
\affiliation{Department of Physics, University of Basel, CH-4056
  Basel, Switzerland}
\author{B.~Trauzettel}
\affiliation{Institute for Theoretical Physics and Astrophysics, 
University of W\"urzburg, D-97074 W\"urzburg, Germany}

\author{C.~Bruder}
\affiliation{Department of Physics, University of Basel, CH-4056 
Basel, Switzerland}
\date{\today}

\begin{abstract}
We investigate a superconducting single-electron transistor
capacitively coupled to a nanomechanical oscillator and focus on the
double Josephson quasiparticle resonance.  The
existence of two coherent Cooper pair tunneling events is shown to
lead to pronounced backaction effects.  Measuring the current and the
shot noise provides a direct way of gaining information
on the state of the oscillator. In addition to an analytical
discussion of the linear-response regime, we discuss and compare
results of higher-order approximation schemes and a fully numerical
solution. We find that cooling of the mechanical resonator is
possible, and that there are driven and bistable oscillator states at low
couplings. Finally, we also discuss the frequency dependence of the
charge noise and the current noise of the superconducting single electron transistor.
\end{abstract}


\maketitle

\section{Introduction}

The cooling of nanomechanical systems by measurement has received a
lot of attention recently. Various procedures like the laser
sideband cooling schemes developed for trapped ions and
atoms,\cite{wineland79} have been proposed as ways to significantly
cool a nanomechanical resonator (NR) coupled to a Cooper-pair 
box,\cite{Martin:2004,Zhang:2005,Hauss:2008,Jaehne:2008} 
a flux qubit,\cite{Wang:2007,You:2008} 
quantum dots,\cite{Wilson-Rae:2004}, trapped ions,\cite{Tian:2004} 
and optical cavities.
\cite{Metzger:2004,Gigan:2006,Arcizet:2006,Kleckner:2006,Schliesser:2006,Corbitt:2007,Thompson:2007,Schliesser:2008,Mancini:1998,Vitali:2002,Nori:2007,Paternostro:2006,Wilson-Rae:2007,Marquardt:2007,Bhattacharya:2007,Genes:2008,Dantan:2007,Kippenberg2007,Marquardt2008,Wineland:2006,Brown:2007,Xue:2007b} 
On the experimental side, optomechanical cooling schemes have been shown
to be promising:
\cite{Metzger:2004,Gigan:2006,Arcizet:2006,Kleckner:2006,Schliesser:2006,Corbitt:2007,Thompson:2007,Schliesser:2008}
the NR was cooled to ultra-low temperatures via either photothermal
forces or radiation pressure by coupling it to a driven cavity.

Another important nanoelectromechanical measurement device which both holds the possibility of very accurate position measurements\cite{lahaye04} as well as of cooling of an NR, is a superconducting single-electron transistor (SSET). Shortly after the theoretical proposals predicting the potential of the
SSET to cool a
nanomechanical system,\cite{blencowe05,clerk05,bennett06}  
this effect has been experimentally
observed.\cite{naik06} Using other detectors for NRs such as normal-state single-electron transistors
\cite{knobel2003aa} or tunnel junctions,\cite{flowers07,balatsky06} it is very difficult to cool the nanomechanical system or drive it into a
non-classical state. These detectors usually act as heat baths with effective temperatures proportional to the transport voltage, which is in practice higher than the bath temperature. The SSET system, on the other hand, shows sharp transport resonances. At those the effective temperature is voltage-independent and can be made very low.

To achieve such challenging goals as ground-state cooling of
NRs,\cite{naik06} or the creation of squeezed oscillator states, a
better understanding of the transport properties of the coupled
SSET-NR system is required. This system is schematically shown in
Fig.~\ref{fig:TheorySSETNEMS}.  Depending on external parameters such
as the gate voltage $V_G$, the bias voltage $V$, and also the
superconducting gap $\Delta$, the SSET supports different types of
resonance conditions. The two most prominent ones are the so-called
Josephson quasiparticle (JQP) and the double Josephson quasiparticle
(DJQP) cycle.\cite{averin89,vandenbrink91}  Whereas the former
involves the coherent tunneling of a Cooper pair at one of the two
junctions followed by a successive tunneling of two quasi-particles at
the other junction, the latter consists of four steps (illustrated in
Fig.~\ref{fig:SSETDJQP} below) that involve a Cooper pair tunneling at
each of the junctions and a quasi-particle tunneling at each of the
junctions. The transport properties of the SSET coupled to an NR close
to the JQP resonance have been analyzed in a recent theoretical
work. \cite{harvey08} Here, we focus on the analysis of the same
coupled quantum system at the DJQP resonance. Since the JQP is a
one-dimensional resonance in the parameter space spanned by $V_G$ and
$V$ and the DJQP is a zero-dimensional resonance in the same parameter
space, all action and backaction effects close to the DJQP resonance
are much more pronounced than close to the JQP resonance. This is of
crucial importance if one wants to manipulate the state of the NR by
measurement of the SSET detector because, in experiments, the typical
coupling between the two quantum systems turns out to be rather weak.

We analyze how the NR can be cooled below the temperature of the
external heat bath and how it can be brought into a (non-thermal)
driven state at the DJQP resonance. Under certain conditions, we find
signatures of bistable solutions of the coupled quantum system of NR
and SSET. It is of particular interest and experimental relevance, to know how
a successful cooling of the NR or the preparation of a driven state
can be observed in transport properties of the SSET such as its
current or current noise. We show that there is a one-to-one
correspondence between interesting state preparations of the NR and
the transport properties of the SSET. This
provides a powerful and feasible tool to initialize and manipulate NR
quantum states by measurement.

The article is organized as follows. In section~\ref{model}, we present
the model for the coupled quantum system of NR and SSET, discuss the
different approximation schemes of the analytical solutions as well as
the calculation scheme behind the exact numerical solution of the
underlying master equation. Then, in section~\ref{oscillator}, we analyze the
oscillator properties by means of the different methods, identifying
interesting quantum states of the NR due to its coupling to the
SSET. Subsequently, in section~\ref{current}, we discuss the current of the SSET
detector and, in section~\ref{noise}, the charge and current noise. It turns out that the
combination of the two transport properties is sufficient to clearly
identify a successful cooling or driven-state preparation of the
oscillator. 
Finally, we present our conclusions in section~\ref{conclusion}. Details of the
calculations are contained in the Appendices.

\section{Model}
\label{model}
\begin{figure}[t]
  \centering
  \includegraphics[width = 0.45 \textwidth]{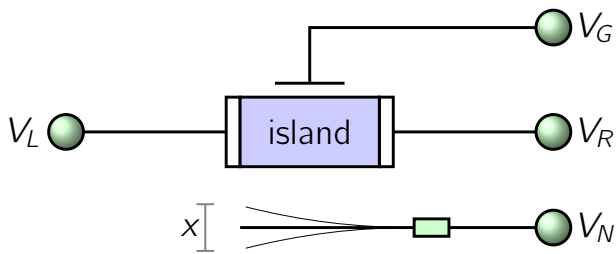}
  \caption{Schematic setup of the SSET-resonator system: Two
    superconducting leads at voltages $V_L$ and $V_R$ are coupled by
    tunnel junctions to a superconducting island. Its chemical potential can
    be tuned by a gate voltage $V_G$. A nearby nanomechanical
    oscillator acts as an $x$-dependent gate.}
  \label{fig:TheorySSETNEMS}
\end{figure}

The system under investigation consists of a superconducting
single-electron transistor (SSET) which is capacitively coupled to a
nanomechanical resonator (NR) as shown schematically in
Fig.~\ref{fig:TheorySSETNEMS}. The total Hamiltonian of the system
reads
\begin{equation}\label{H}
 H = H_L + H_R + H_I + H_T + H_C + H_N + H_{N,I} \;.
\end{equation} 

The first three terms $H_{L,R,I}$ are standard BCS Hamiltonians and
describe two superconducting leads (left and right) and a
superconducting island,
\begin{equation}
 H_\alpha = \sum_{k,\sigma} \epsilon_{\alpha k \sigma} c^\dag_{\alpha k \sigma} c_{\alpha k \sigma}^{\phantom{\dagger}}\;.
\end{equation} 
Here, $c_{\alpha k \sigma}$ are annihilation operators for
quasiparticles of momentum $k$ and spin $\sigma$ in the system
$\alpha$ ($\alpha = L,R,I$). The dispersion relation $\epsilon_{\alpha
  k \sigma}$ accounts for the superconducting gap of width $2 \Delta$
which we assume to be equal for the three systems. The chemical
potentials in the left and right leads are determined by the applied
bias voltage $V = V_L - V_R$, while the island chemical
potential can be tuned by applying a gate voltage $V_G$ (see
Fig.~\ref{fig:TheorySSETNEMS}).

The left and right leads are connected to the central island by
quasiparticle tunneling and Cooper pair tunneling. Denoting by
$\phi_\alpha$ the superconducting phase difference at the junction
$\alpha = L, R$, we use the following quasiparticle tunneling term
\begin{equation}
 H_{T,qp} = \sum_{\alpha=L,R} e^{-i\phi_\alpha/2} \sum_{k,q,\sigma} 
T^{\phantom{\dag}}_{kq} c^\dag_{\alpha k \sigma} c_{I q \sigma}^{\phantom{\dag}} + \hc
\;,
\end{equation} 
where $T_{kq}$ are the tunneling amplitudes which can be used to
calculate \cite{choi01} the quasiparticle tunneling rates
$\Gamma_{L,R}$. Cooper pair tunneling is accounted for by the term
\begin{equation}
 H_{T,CP} = - \sum_{\alpha = L,R} J_\alpha \cos \phi_\alpha\;,
\end{equation} 
where $J_\alpha$ are the Josephson energies of the two
junctions. Hence, the total tunneling Hamiltonian is given by $H_T =
H_{T,qp} + H_{T,CP}$.

The final ingredient for the SSET Hamiltonian is the Coulomb energy of
the island. If we denote by $n_L$ and $n_R$ the number of electrons
that have tunneled from the island to the left and right lead,
respectively, then $n = -n_L - n_R$ is the excess number of electrons
on the island. The charging term can be written as
\begin{equation}
 H_C = E_C(n + n_0)^2 + e V n_R\;,
\end{equation} 
where $E_C$ is the charging energy and $n_0$ can be controlled by the
gate voltage (see Appendix~\ref{app:coup}). In terms of the
capacitances of the two junctions $C_{L,R}$, the gate $C_G$ and the
resonator $C_N$, the charging energy is given by $E_C = e^2/ (2
C_\Sigma)$, where $C_\Sigma = C_L + C_R + C_G + C_N$ is the total
capacitance.

Next, we focus on the coupling of the SSET to the NR. The latter can
be regarded as a harmonic oscillator of frequency $\Omega$ and mass
$M$ and is therefore described by the Hamiltonian
\begin{equation}
 H_N = \hbar \Omega\left(\nosc + \tfrac{1}{2} \right) =
 \frac{p^2}{2M} + \tfrac{1}{2} M \Omega^2 x^2\;.
\end{equation} 
The NR is held on a constant voltage $V_N$ and hence acts on the SSET
as an additional gate with an $x$-dependent capacitance
$C_N(x)$. Therefore, the presence of the NR modifies the charging term
$H_C$. Expanding the contribution for small displacements $x$ and
retaining only the lowest order, one finds that the coupling between
SSET and NR is given by
\begin{equation}\label{HNI}
 H_{N,I} = - A\, n\, x\;,
\end{equation} 
where the coupling constant $A$ depends in a non-trivial way on the
voltages and capacitances of the system and can be regarded as an
effective parameter (see Appendix~\ref{app:coup}). Note that this
expansion is only valid for displacements $x$ which are small compared
to the distance $d$ between the SSET and the NR, i.e.~$x/d \ll 1$. Upon
continuing the expansion, one encounters terms proportional to $n^2$
and to $x^2$ which will be neglected here.

Due to the complex structure of the full Hamiltonian (\ref{H}) one
should not hope for an exact solution in all regimes. Instead, we will
make several assumptions which will enable us to investigate the
transport properties of this system at a particular point in the
parameter space.

First, we will briefly review the transport properties of the bare
SSET without coupling to the NR. While the capacitances, Josephson
energies and quasiparticle tunneling rates are essentially determined
by the experimental setup, the most important tunable parameters are
the bias voltage $V$ and the gate voltage $V_G$. The transport
properties of the SSET are then determined by how these voltages are
related to the superconducting gap $2\Delta$ and the charging energy
$E_C$.

For high bias voltages $e V > 4 \Delta$, the difference in chemical
potentials allows quasiparticles on both junctions to overcome the
superconducting gap and a quasiparticle current can flow. But even for
lower bias voltages, one observes a finite current at certain values
of the gate voltage. A possible mechanism is the
Josephson-quasiparticle (JQP) resonance which is a cyclic process that
starts with the tunneling of a Cooper pair on one of the junctions
followed by two subsequent quasiparticle tunneling events on the other
junction.\cite{averin89,nakamura96} This process is possible above a
lower bias voltage threshold, $e V > 2 \Delta + E_C$.

\begin{figure}[t]
  \centering
  \includegraphics[width = 0.45 \textwidth]{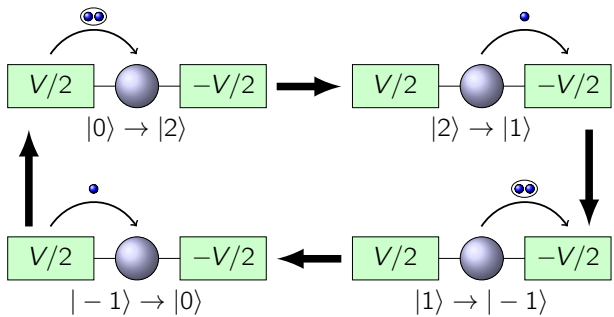}
  \caption{Illustration of the DJQP cycle: (i) Cooper pair tunneling
    through the left junction, (ii) quasiparticle tunneling through
    the right junction, (iii) Cooper pair tunneling through the right
    junction and (iv) quasiparticle tunneling through the left
    junction.}
  \label{fig:SSETDJQP}
\end{figure}

For even lower bias voltages, isolated current resonances can be
observed which are due to the onset of the double Josephson
quasiparticle (DJQP) resonance. A schematic picture of this process is
shown in Fig.~\ref{fig:SSETDJQP}. It starts with a Cooper pair
tunneling across, say, the left junction. Next, a quasiparticle
tunnels out through the right junction, followed by a Cooper
pair. Finally, after a quasiparticle tunnels through the left
junction, the initial system state is reached again.

This process is energetically allowed only in a restricted parameter
regime: Cooper pair tunneling is only possible if the chemical
potentials of the lead and the island (taking into account the Coulomb
energy) are on resonance while quasiparticle tunneling requires a
difference in chemical potentials sufficient to overcome the
superconducting gap. For the DJQP process, it is easy to show that the
resonances occur at bias voltages $e V = 2 E_C$ and half-integer
island charges $n_0$.

The parameter regime which we investigate is therefore characterized
by a charging energy $E_C$, a superconducting gap $2 \Delta$ and a
bias voltage $V$ which are of the same order of magnitude. Roughly
speaking, these energy scales are very large compared to the
quasiparticle tunneling rates $\Gamma_{L,R}$, the Josephson energies
$J_{L,R}$ and the oscillator energy $\Omega$.

\subsection{Derivation of a Liouville equation}

Due to the small tunneling rates, only sequential tunneling will
contribute to the transport whereas higher-order (cotunneling)
processes are suppressed. This suggests describing the system by a
master equation in the Born-Markov approximation.

For this purpose, we treat the BCS Hamiltonians $H_L + H_R + H_I$ as a
fermionic bath for the remaining system. Then, system and bath are
only coupled by the quasiparticle Hamiltonian $H_{T,qp}$. Using the
Born approximation corresponds to disregarding cotunneling processes
while the Markov approximation is valid as long as there is a
separation of time scales between the system and the bath degrees of
freedom. Introducing the system and bath Hamiltonians
\begin{eqnarray}
 H_S & = & H_C + H_{T,CP} + H_N + H_{N,I}\;, \\
 H_B & = & H_L + H_R + H_I\; ,
\end{eqnarray} 
and using the Born-Markov approximation leads to the following
master equation for the reduced density matrix $\rho(t)$ of the
system,
\begin{eqnarray}\label{eq:master}
 \dot{\rho}(t) & = &\mathcal{L} \rho(t) \\
& = & -\frac{i}{\hbar} \left[H_S, \rho(t)\right] \nonumber \\
& - &
 \frac{1}{\hbar^2} \int_0^\infty d\tau \Tr_B
 \left[ H_{T,qp}, \left[ H_{T,qp}(-\tau), \rho(t) \otimes \rho_B
 \right]\right] \;, \nonumber 
\end{eqnarray} 
where $\rho_B$ is the bath density matrix. The time dependence of the
$H_{T,qp}$ operator is governed by the Hamiltonian $H_S + H_B$. The
density matrix $\rho$ contains information only about the charge and
the oscillator degrees of freedom and can, for example, be written in
the basis $\ket{n,n_R,x}$ of island charge states $\ket{n}$, the
amount of charge $\ket{n_R}$ which has tunneled through the right
junction, and the oscillator coordinate $\ket{x}$. This approach
allows the calculation of the transport properties of the system via
charge counting.\cite{clerk03,armour04} In order to investigate the
transport at the DJQP resonance, it is sufficient to consider a finite
number of basis states for the island charge $n$. As a single DJQP
cycle involves four charge states, we can restrict the basis to the
states $\ket{-1}$, $\ket{0}$, $\ket{1}$, and $\ket{2}$ which significantly
reduces the complexity of the problem, since it is thus sufficient
to study a reduced density matrix as described in section~\ref{liouville}.
This choice of charge states corresponds to $n_0 = - 1/2$.

As long as one is only interested in oscillator properties or the current through the SSET, the $\ket{n_R}$ states can be traced out, and an effective master equation acting on the Hilbert space of island charge and oscillator position, spanned by the states $\ket{n, x}$, can be obtained. On the other hand, for the calculation of the current noise, the $\ket{n_R}$ degree of freedom has to be taken into account explicitly, as will be explained later on. For now, we proceed with the $n_R$-independent case.

As in the case of the JQP,\cite{harvey08} the Liouvillian obtained
from Eq.~(\ref{eq:master}) can be written as a sum of three
contributions
\begin{align} \label{eq:liouville 3 contributions}
	\mathcal{L} = \mathcal{L}_{H_S} + \mathcal{L}_{qp} + \mathcal{L}_{CL}\;,
\end{align}
where $\mathcal{L}_{H_S}$ governs the coherent evolution of the
system, $\mathcal{L}_{qp}$ is a dissipative term due to quasiparticle
tunneling and $\mathcal{L}_{CL}$ is a Caldeira-Leggett type contribution
introduced to model the coupling of the harmonic oscillator to a
finite-temperature environment. Explicitly,
\begin{align}\label{eq:liouville details}
	\mathcal{L}_{H_S} \rho
&=
	-\frac{i}{\hbar} [H_S,\rho] \;,\\
	\mathcal{L}_{qp}  \rho
&= 
	\Gamma_L \hat p _{-1,0} \rho \hat p _{-1,0}^\dagger 
      - \frac{1}{2} \Gamma_L \left \{ \hat p_{-1,-1},\rho  \right\} \\
&+ 	\Gamma_R \hat p_{2,1} \rho \hat p _{2,1}^\dagger 
      - \frac{1}{2} \Gamma_R \left \{ \hat p_{2,2} , \rho \right \}
\;, \nonumber \\
	\mathcal{L}_{CL} \rho 
&=
	- \frac{D}{\hbar^2} [x,[x, \rho]] - \frac{i \gamma_{\mathrm{ext}} M}{2 \hbar} [x,\{v, \rho\}] \;,
\end{align}
where $\hat p_{kj} = \ket{j} \bra{k}$ acts
on the charge states of the island and is utilized here to 
describe the quasiparticle tunneling event that changes the charge state 
from $\ket{k}$ to $\ket{j}$. The energy dependence of the
quasiparticle tunneling rates $\Gamma_L$ and $\Gamma_R$ is weak and
will therefore be neglected in the following. The diffusion constant
$D$ and external damping rate $\gamma_{\mathrm{ext}}$ are related via
a fluctuation-dissipation relation
\begin{align}
	  D &=  M \gext \frac{\hbar \Omega}{2} 
	        \coth\left( \frac{\hbar \Omega}{2 k_B \TB}\right) \notag \\
	    &=  M \gext\ \hbar \Omega 
	        \left( \expcts{\nosc}{} + \frac{1}{2} \right)\;.
\end{align}
For $\hbar \Omega \ll k_B T_B$, the diffusion constant can be
approximated by $D = M \gext k_B \TB$.

In the general case, it is neither possible to calculate exactly 
the steady-state properties nor the transport properties of the coupled
system using the Liouville superoperator 
(\ref{eq:liouville 3 contributions}).  Thus, approximation schemes must be
employed. In the following two subsections, we describe in detail the
two complementary approximation schemes we use to study the coupled
SSET-oscillator system.

\subsection{Mean-field approach}
\label{subsec:mean_field_approach}

Physical quantities can be calculated by evaluating the matrix
elements of the density matrix $\rho(t)$. It turns out that 
oscillator properties and the average current can be written in terms of expectation values of the form
\begin{equation}\label{defrho}
 \expct{x^n v^m \hat p_{kj}}{} = 
\Tr_\osc  \sum_{n_R} \bra{k, n_R} x^n v^m \rho(t) \ket{j, n_R + k -j} \;,
\end{equation} 
where $x$ and $v$ are the position and velocity operators of the oscillator and $\hat p_{kj} = \ket{j} \bra{k}$. The trace over the oscillator degrees of freedom $\Tr_\osc$ will be used in the position basis $\Tr_\osc (\cdot) = \int dx \bra{x} \cdot \ket{x}$ as well as in the phonon number basis where $\Tr_\osc (\cdot) = \sum_{n_\osc=0}^\infty \bra{n_\osc} \cdot \ket{n_\osc}$. For the uncoupled SSET tuned closely to the DJQP resonance, the
average current can be calculated straightforwardly, as taking
the matrix elements of the master equation (\ref{eq:master}) leads to
a closed set of equations.\cite{clerk03} If the NR is included,
however, the coupling terms will lead to equations involving matrix
elements of the form $\expcts{x \hat p_{kj}}{}$. Calculating their time
evolution leads to ever higher-order terms of the form $\expcts{x^n
  v^m \hat p_{kj}}{}$, so that the set of differential equations never
closes. Hence, a truncation scheme is needed. A standard route is to
truncate the system of equations by assuming a vanishing $n$th-order
cumulant $\cumu{x^n \hat p_{kj}}{}$. This allows one to rewrite 
$n$th-order expectation values in terms of expectation values of order $n-1$
and hence to arrive at a closed, albeit non-linear, set of equations.

We use and compare these approximations for $n=1$ 
(which we call \emph{thermal-oscillator approximation}) and $n=2$ 
(\emph{Gaussian approximation}). These two levels of approximation 
are related to what was called `mean 1' and `mean 2' in 
Ref.~[\onlinecite{harvey08}].  In order to give
an estimate of the physical quality of the truncation scheme, we also
compare our results to exact numerical calculations.

Whereas the expectation values of the form (\ref{defrho}) are sufficient
for the calculation of the oscillator properties and the SSET current,
the calculation of the noise requires a slightly extended approach. In
order to keep track of the transfered charge, one has to investigate
the dynamics of $n_R$-resolved expectation values. It turns out (see Appendix~\ref{app:noise}) 
that the noise can be rewritten in terms of expectation values of the
operators 
\begin{equation}
\hat{p}_{kj}^{n_R'n_R} = \ket{j}\bra{k} \otimes \ket{n_R}\bra{n_R'}\;.
\end{equation} 
Note that only elements which conserve the number of charges,
$j + n_R = k + n_R'$, are finite.
An analogous truncation scheme can be applied to expectation values
containing these operators. Similar approaches have been
used extensively to describe nanoelectromechanical
systems.\cite{armour04,armour04b,rodrigues05,rodrigues05b,blencowe05,doiron06,harvey08}

\subsection{Numerical solution of the Liouville equation}
\label{liouville}
To complement the analytical mean-field approach described in the last
subsection, we also use a numerical approach to study the properties
of the NR coupled to an SSET near the DJQP resonance. First, we present the approach taken for the calculation of the current and the oscillator properties, where the $n_R$ degree of freedom plays no role. 
Subsequently, we will demonstrate how to extend this approach for the noise calculation, where $n_R$ has to be taken into account.

To calculate the current and the oscillator properties, we write the density matrix in the $\lvert n, \nosc \rangle$ basis, with
$\nosc$ being the phonon quantum number of the oscillator and $n$ the charge of the SSET. The spectrum of the harmonic oscillator is naturally not finite, so we
need to truncate it and consider only its $N_{max}$ lowest energy
eigenstates.
To describe the DJQP cycle, the reduced density matrix $\rho$ is of
dimension $(4 N_{max} \times 4 N_{max})$.

The Liouville superoperator $\mathcal{L}$ is a two-sided operator, in
the sense that it acts both from the left and the right of the density
matrix [cf.~Eq.~(\ref{eq:liouville details})]. It can be transformed
to a single-sided operator using a property of the matrix
vectorization operation: the vectorized form of a product of three ($4N_{max}
\times 4 N_{max}$) matrices \textbf{A},\textbf{B},\textbf{C} can be written as
a single product of an ($16 N_{max}^2 \times 16 N_{max}^2$) matrix with an ($16N_{max}^2 \times 1$)
vector via the relation $\mathrm{vec}( \textbf{ABC}) = (\textbf{C}^T
\otimes \textbf{A}) \mathrm{vec}(\textbf{B})$, where the superscript $T$ denotes the matrix transposition and $\otimes$ a Kronecker product.\footnote{An element $D_{\alpha \beta}$ of the matrix $D =(\textbf{C}^T \otimes \textbf{A})$, with $\textbf{A},\textbf{C}$ two $(4N_{max} \times 4 N_{max})$ matrices, is therefore given by $D_{\alpha \beta} = C_{j i} A_{k l}$ with $\alpha = 4N_{max}(i-1)+k$ and $\beta= 4N_{max}(j-1)+l$} The matrix
representation of the Liouville superoperator is therefore of order ($16 N_{max}^2 \times 16 N_{max}^2$).

To illustrate how the aforementioned vector identity can be used, we apply it to the coherent evolution contribution to the Liouville equation [Eq.~\eqref{eq:liouville details}]. In this case, we find
\begin{align}
\mathcal{L}_{H_S} \rho &= -\frac{i}{\hbar} [H_S,\rho] \notag, \\
\to \ \mathrm{vec}(\mathcal{L}_{H_S} \rho) &= -\frac{i}{\hbar} \left( \mathbf{I} \otimes H_S +H_S^T \otimes \mathbf{I} \right) \mathrm{vec}( \rho),
\end{align}
where the matrix representation of the identity matrix  $\mathbf{I}$ and of the system Hamiltonian $H_S$ is of order $(4N_{max} \times 4 N_{max})$.

To find the vectorized form of the stationary density matrix
$\rho_{stat}$, defined from $\mathcal{L}\rho_{stat} = 0$, we calculate
the null-space of the Liouville matrix. Using the normalization condition
$\Tr [\rho_{stat}] = 1$, the stationary density matrix $\rho_{stat}$
can be determined uniquely. The bad scaling
$\mathcal{O}(N_{max}^4)$ of the Liouvillian size with the truncation
point in the oscillator spectrum makes the numerical eigenvalue problem
very challenging. Luckily, in this problem the Liouville matrix
displays a high sparsity degree, and sparse eigensolvers can be
used. Our implementation uses the shift-invert mode of the
ARPACK \cite{lehoucq96} eigensolver in combination with the
PARDISO \cite{schenk04,schenk06} linear solver to compute the first few
($\sim 5$) eigenvalues of $\mathcal{L}$ with the lowest magnitude
($\lambda_0$) as well as the associated eigenvectors. The calculated
magnitude of the smallest eigenvalue can be used to verify the validity
of the truncation scheme: when enough Fock states are kept 
we find $|\lambda_0| < 10^{-15}$ which is below the desired
precision limit of $10^{-10}$.

To improve the speed of the calculation and, more importantly, to
increase numerical accuracy, we do not need to explicitly solve for
those matrix elements of $\rho_{stat}$ which, due to the considered
Hamiltonian, have to be zero. For example, coherence can only be
created between two charge states $\lvert k \rangle$ and $\lvert j
\rangle$ if $|k-j| = 2$, since only these pairs of states are coupled
by Josephson tunneling. Therefore, all density-matrix elements
$\bra{k} \rho \ket{j}$ where $|k-j|$ is odd are zero. Using this argument, the
size of the Liouville matrix can be reduced to $(8N^2_{max} \times
8N^2_{max})$.

The use of sparse solvers also minimizes the required memory for the
calculation of the eigenvalues, allowing problems of relatively large
size ($N_{max} \lesssim 150$) to be solved on a desktop
computer. Also, we note that, contrary to what was discussed in
Ref.~[\onlinecite{flindt04}], no manual preconditioning was needed to
achieve high numerical accuracy. To allow for the numerical approach
to be used in the driving regime, where we expect the average energy
of the oscillator to be relatively high, we had to make a
supplementary approximation. In this case, we assumed that coherence could 
develop only between states of the oscillator that are not too far
away in energy from each other, setting $\langle n_{osc} | \rho  |
n_{osc}^\prime \rangle = 0$ for $|n_{osc}-n_{osc}^\prime| \gtrsim 60 $. This allowed for
$N_{max}$ to be set as high as 750 on a standard
workstation. Moreover, in the cases where it was possible to compare
directly the results of the calculation with and without this last
approximation, we found that they were identical (within our numerical
accuracy). 

While this approach is viable for the calculation of the oscillator properties and the current, it fails to keep track of the tunneled charge $n_R$ and thus cannot be used to calculate the current noise. A straightforward inclusion of the $\ket{n_R}$ states is numerically impossible as the corresponding Hilbert space is of infinite dimension. However, this problem can be circumvented by considering the $n_R$-resolved density matrices $\rho^{(n_R)}$ ($n_R \in \mathbb{Z}$), which are submatrices of the complete density matrix $\rho$, whose entries are defined by
\begin{align}
 \bra{k} \rho^{(n_R)} \ket{j} = \bra{k, n_R'} \rho \ket{j, n_R}\; ,
\end{align}
where the relation between $n_R$ and $n_R'$ is given by charge 
conservation. At the DJQP resonance, we have $n_R' = n_R - 2$ for $(k,j) = (1,-1)$, $n_R' = n_R + 2$ for $(k,j) = (-1,1)$ are $n_R' = n_R$ otherwise.\cite{clerk03}
Note that we did not write out the oscillator degree of freedom explicitly in this matrix. Calculating the time evolution of these matrix elements according to Eq.~\eqref{eq:master}, one finds
\begin{align}\label{drhototal}
 \frac{d}{dt} \rho^{(n_R)} 
&=
 \left[\mathcal{L} - \mathcal{I}_{qp} - \mathcal{I}_{CP}^+ - \mathcal{I}_{CP}^- \right] \rho^{(n_R)}  \\
&+ \mathcal{I}_{qp} \rho^{(n_R-1)}
 + \mathcal{I}_{CP}^+ \rho^{(n_R+2)} + \mathcal{I}_{CP}^- \rho^{(n_R-2)}
 \ .\notag
\end{align}
As expected, the tunneling leads to a coupling between density matrices of different $n_R$. It is produced by the current superoperators describing the quasiparticle and the Cooper pair tunneling, which are defined as
\begin{align}
 \mathcal{I}_{qp} \rho 
&= \Gamma_R \ket{1}\bra{2} \rho \ket{2} \bra{1}\; ,
 \\
 \mathcal{I}_{CP}^+ \rho 
&= 
 - \frac{i J_R}{2 \hbar} \left[ \ket{1} \bra{-1} \rho \ket{-1} \bra{-1}
 + \ket{1} \bra{-1} \rho \ket{1} \bra{1} \right] \notag \; , \\
 \mathcal{I}_{CP}^- \rho 
&= 
 - \frac{i J_R}{2 \hbar} \left[ \ket{-1} \bra{1} \rho \ket{-1} \bra{-1} 
 + \ket{-1} \bra{1} \rho \ket{1} \bra{1} \right] \notag \ .
\end{align}
It is important to realize that by writing the Liouville equation in terms of $n_R$-resolved density matrices and current superoperators, we have achieved a description of the system in terms of only the $\ket{n, n_\osc}$ states again. This, however, comes at the price of having to deal with an infinite number of density matrices, $\rho^{(n_R)}$. Still, following the approach of Ref.~[\onlinecite{flindt05}] it will turn out that convenient expressions for the current and the noise can be formulated in terms of these current superoperators.

\section{Oscillator properties}
\label{oscillator}
As mentioned before we treat the NR as a harmonic oscillator and we
use the master equation to investigate the time evolution of the
mean displacement
\begin{align}
  \expct{x}{} = \Tr_{n_R} \Tr_n \Tr_{\osc} \left[ \rho(t) x \right]\;,
\end{align}
and of the velocity $v$, correspondingly. Here, $\Tr_n (\cdot) = \sum_{n=-1}^2 \bra{n} \cdot \ket{n}$ denotes the trace over the island charge, while $\Tr_{n_R} (\cdot) = \sum_{n_R=-\infty}^{\infty} \bra{n_R} \cdot \ket{n_R}$ traces over the tunneled charge. Likewise, the master
equation will allow us to calculate expectation values of higher order
like $\expcts{x^2}{}$ and $\expcts{v^2}{}$ which are required for the
calculation of the oscillator energy.

For a linear coupling of the NR to the SSET as in Eq.~(\ref{HNI}), we
find the following equations describing the time evolution of the
oscillator coupled to the SSET
\begin{align}
  \frac{d}{dt} \expct{x}{} &= \expct{v}{}\;, \\
  \frac{d}{dt} \expct{v}{} &= - \Omega^2 \expct{x}{}
                              - \gext \expct{v}{}
                              + \frac{A}{M} \expct{n}{}\;,
\end{align}
where $\expcts{n}{} = \sum_k k
\expct{\hat p_{kk}}{}$ is the expectation value of the island occupation
and $\gext$ accounts for the external damping. The stationary limit,
where $\tfrac{d}{dt} \expcts{v}{} = \tfrac{d}{dt} \expcts{x}{} = 0$,
can be regarded as the long-time limit when the oscillatory behavior
has been damped by the thermal bath and thus $\expcts{v}{} = 0$ and
$\expcts{x}{} = (A / M \Omega^2) \expcts{n}{}$.  If the coupling $A$ to
the SSET is zero, the oscillator stays in its equilibrium
position at $\expcts{x}{} = 0$.  For finite coupling, due to the
electromagnetic repulsion the NR equilibrates in a position
$\expcts{x}{} \not= 0$ proportional to the coupling and the charge
$\expcts{n}{}$ on the island.  Note that the influence of the SSET on the NR
is of the first order in the coupling $A$. This regime has already
been studied in some detail \cite{blencowe05, clerk05} and it was shown
that the SSET acts as an effective thermal bath for the NR. As we will
illustrate further on, the signature of the NR in the transport properties of the SSET is
of second order in the coupling and is clearly visible in the
current and the noise properties of the SSET.

To study the influence of the NR on the SSET we introduce
dimensionless quantities which are normalized to motional quanta of
the oscillator. Using the frequency $\Omega$ and the harmonic
oscillator length,
\begin{align}
  x_0 &= \sqrt{\frac{\hbar}{2 M \Omega}}\;,
\end{align}
as units we define $\ti{x} = x/x_0$, $\ti{t} = \Omega t$ and $\ti{v} =
v/ \Omega x_0$, i.e.~we normalize all variables with respect to
``oscillator'' quantities. This allows an easier comparison with the
experiment where, for example, the bias voltage can be varied at
constant coupling. Consequently, the equations of motion can be
rewritten as
\begin{align}
  \frac{d}{d\ti{t}} \expct{\tx}{} &= \expct{\tv}{}, \\
  \frac{d}{d\ti{t}} \expct{\tv}{} &= - \expct{\tx}{}
                              - \tgext \expct{\tv}{}
                              + 2 \ti{A} \expct{n}{}\;.
\end{align}
where $\tgext = \gext/\Omega$ and $\ti{A} = x_0 A/\hbar \Omega$.

Not only is the equilibrium position of the resonator shifted by the
coupling to the SSET, but also the cumulants of the position and
velocity of the NR, i.e.~$\cumu{x^2}{} = \expcts{x^2}{} -
\expcts{x}{}^2$, are influenced by this coupling:
\begin{align}
  \frac{d}{d\ti{t}} \cumu{\tx^2}{} &= \cumu{\{\tx, \tv\}_+}{}\;, \\
  \frac{d}{d\ti{t}} \cumu{\{\tx, \tv\}_+}{} 
   &= 2 \cumu{\tv^2}{} - 2 \cumu{\tx^2}{}
    - \tgext \cumu{\{\tx, \tv\}_+}{}
   \notag \\ &
    + 4 \ti{A} \left( \expct{n\; \tx}{} - \expct{n}{} \expct{\tx}{} \right)\;, \\
  \frac{d}{d\ti{t}} \cumu{\tv^2}{} &= 
    - \cumu{\{\tx, \tv\}_+}{}
    - 2 \tgext \cumu{\tv^2}{}
    + 4 \tgext \ti{T}_B
   \notag \\ &
    + 4 \ti{A} \left( \expct{n\; \tv}{} - \expct{n}{} \expct{\tv}{} \right)\;,
\end{align}
where $\ti{T}_B = k_B \TB/\hbar \Omega$ and $\{\cdot,\cdot\}_+$
denotes the anti-commutator. In the stationary limit this leads to
\begin{align}
  \cumu{\{\tx, \tv\}_+}{} &= 0\;, \\
  \cumu{\tv^2}{} &= 2 \ti{T_B} + 2 \ti{A} \left( \expct{n\; \tv}{} 
                           - \expct{n}{} \expct{\tv}{} \right)/\tgext\;,
\label{eq:v2} \\  
  \cumu{\tx^2}{} &= \cumu{\tv^2}{} + 2 \ti{A} \left( \expct{n\; \tx}{}
                           - \expct{n}{} \expct{\tx}{} \right)\;.
\label{eq:x2}
\end{align}

To lowest (linear) order, which we refer to as the 
\textit{thermal-oscillator approximation}, we assume that $\expcts{n v}{} =
\expcts{n}{} \expcts{v}{}$. This is identical to assuming that the
correlations between $n$ and $v$ vanish, i.e.~$\cumu{n\, v}{} =
0$. Consequently, the fluctuations of the harmonic oscillator are not
influenced by the SSET such that the virial theorem $\cumu{\tv^2}{} =
\cumu{\tx^2}{}$ and the equipartition theorem $\cumu{\tv^2}{} = 2\,
\ti{T}_B$ are fulfilled in the high-temperature limit $\ti{T}_B \gg
1$. The resonator is thus in a \textit{thermal state} determined only
by $\ti{T}_B$ and $\tgext$. In the thermal-oscillator approximation 
analytic expressions for the current and noise in the
SSET can be derived and will be discussed in the upcoming sections.

The thermal-oscillator approximation is justified for weak
coupling between the SSET and the NR, but fails for stronger coupling,
since the observables of the oscillator become entangled with the
charge state of the SSET. As was already observed
before,\cite{clerk05, blencowe05} an increased coupling can drive the
oscillator to a non-thermal state characterized by a finite
$\expcts{n v}{} - \expcts{n}{} \expcts{v}{} \not= 0$, where the
virial and equipartition theorems no longer hold.

In order to investigate this regime, we have to go to the next order
in our approximation which means taking the fluctuations of
$\cumu{nx}{}$ into account, but assuming all higher-order cumulants to
vanish, e.g.~$\cumu{n x^2}{} = 0$. This will be referred to as the
\textit{Gaussian approximation} since for a Gaussian distribution all
cumulants $\cumu{x^n}{}$ for $n > 2$ are zero and the resonator is
fully described by the two lowest moments. Under this assumption, we
can express expectation values of the form $\expcts{x^2 \hat p_{kj}}{}$
as \emph{products} of the lower-order expectation values $\expcts{x
  \hat p_{kj}}{}$, $\expcts{x^2}{}$, $\expcts{x}{}$ and
$\expcts{\hat p_{kj}}{}$. While this approach leads to a closed set of
differential equations, the set will now be non-linear and has to be
solved numerically.

In principle, this approximation scheme can be continued to even
higher orders.\cite{harvey08} However, since the Gaussian
approximation works well for the low-coupling regime we are interested
in, we do not go beyond it. Ultimately, for even stronger coupling,
the linear coupling between the SSET and the NR itself becomes
questionable.

In order to investigate the oscillator state in more detail, we study
the energy $E = \tfrac{1}{2} M \Omega^2 x^2 + \tfrac{1}{2} M v^2$, in
dimensionless quantities
\begin{align}
  \expct{\frac{E}{\hbar \Omega}}{}
    &= \frac{1}{4} \left( \expct{\tx^2}{} + \expct{\tv^2}{} \right)
     = \expct{n_\mathrm{osc}}{} + \frac{1}{2}\;.
\end{align}
Previous work \cite{clerk05,rodrigues05} has focused on the
fluctuations of the number of charges on the island, $n$, which in
linear response can be described by an effective damping and effective
temperature. We will discuss this approach in more detail in the
context of the charge noise. For an identification of the oscillator
state, though, we choose a different route and investigate the energy
of the NR. In the stationary limit, using
Eqs.~\eqref{eq:v2} and \eqref{eq:x2}, we find for the energy
\begin{align}
  \expct{\frac{E}{\hbar \Omega}}{}
&= \ti{T}_B +  \ti{A}\ \frac{2 \cumu{n\tv}{} + \tgext \cumu{n\tx}{}}{2 \tgext}
+ \frac{1}{4} \expct{\tx}{}^2  \;.
\end{align}
A finite $\expcts{x}{} \not= 0$ provides additional potential energy,
but the contribution is small, since $\expcts{\tx}{}^2 = 4\ti{A}^2
\expcts{n}{}^2$. Therefore, it is the correlations of the entangled
SSET-NR system contained in the second term, which have the potential
to drive the system out of a thermal state.

\begin{figure}[t]
  \centering
  \includegraphics[width = 0.45 \textwidth]{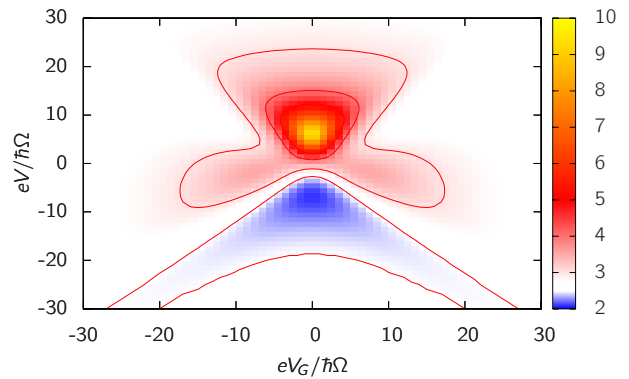}
  \caption{Oscillator energy in units of $\hbar \Omega$ in the
    Gaussian approximation as a function of the gate voltage
    $e V_G/\hbar \Omega$ and the bias voltage $eV/\hbar \Omega$ where $(0,0)$
    denotes the resonance. The parameters used are
    $\ti{\Gamma}_{L} = \ti{\Gamma}_R = 10, 
    \ti{J}_{L} = \ti{J}_R = 2, \tgext = 10^{-4}$, $\tTB = 2.5$ and
    $\tA = 0.02$. In the red-detuned area ($V < 0$, $- V < V_G < V$),
    cooling below the bath temperature is visible (blue
    region). Driving can be observed in the blue-detuned case ($V >
    0$, $- V < V_G < V$). The highest energies are obtained in the
    yellow region.}
  \label{fig:StabilityV1}
\end{figure}

\begin{figure}[t]
  \centering
  \includegraphics[width = 0.45 \textwidth]{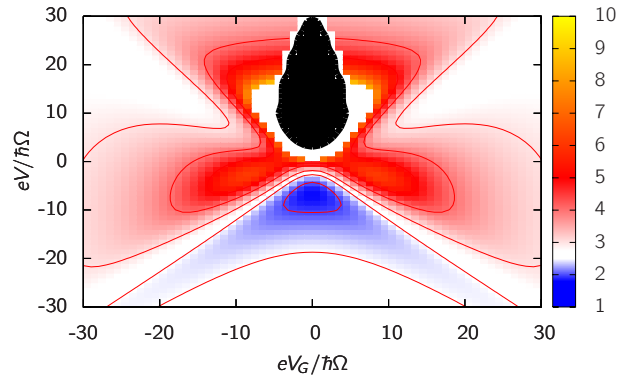}
  \caption{Oscillator energy in the Gaussian approximation
    in units of $\hbar \Omega$ for the
    Gaussian approximation as a function of the gate voltage
    $eV_G/\hbar \Omega$ and the bias voltage $eV/\hbar\Omega$ 
    for increased coupling $\tA = 0.03$. 
    Cooling and driving effects are increased as compared to
    Fig.~\ref{fig:StabilityV1}. In the black area two stable and one
    unstable solution are found, i.e.~the system becomes bistable. The
    area grows for stronger coupling.}
  \label{fig:StabilityV3}
\end{figure}

The results for a calculation of the energy in the Gaussian
approximation for a typical, experimentally relevant set of parameters
are shown in Figs.~\ref{fig:StabilityV1} and \ref{fig:StabilityV3}
where we display the oscillator energy as a function of gate voltage
$V_G$ and the bias voltage $V$ measured away from the resonance
position. As the DJQP cycle contains two Cooper pair tunneling
events, there are two resonance conditions which have to be met and
which can be controlled by adjusting the bias and gate voltages.

The physical picture can be explained most clearly if we assume $V_G =
0$, which corresponds to a vertical cut in
Fig.~\ref{fig:StabilityV1}. If the system is blue-detuned from a
resonance ($V > 0$), the tunneling Cooper pairs transfer a part of
their energy to the oscillator in order to be able to tunnel. This
leads to driving of the oscillator. On the contrary, for a red-detuned
resonance ($V < 0$), the Cooper pairs can absorb energy from the
oscillator, leading to cooling. A similar result was already found
using a linear-response approach in Ref.~[\onlinecite{clerk05}].

In the regime where both resonances involved in the DJQP cycle are
blue-detuned ($V > 0$, $-V < V_G < V$), we find a particularly strong
driving of the oscillator. In the white regions of
Fig.~\ref{fig:StabilityV3}, energies of the
order $10^3 \hbar \Omega$ (depending on the system parameters) are
reached even for rather small coupling of the order $\ti{A} \approx
0.02$. The numerical solution of the Liouville equation reveals
moreover that the resulting oscillator state is highly non-thermal,
i.e.~the distribution function of oscillator states $P(\nosc)$
strongly deviates from a Boltzmann distribution.

\begin{figure}[t]
  \centering
  \includegraphics[width = 0.45 \textwidth]{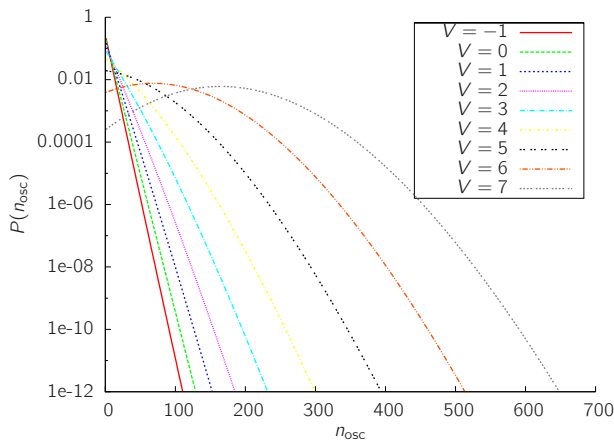}
  \caption{Distribution $P(\nosc)$ of the oscillator phonon number for different values of the bias voltage
    $eV/\hbar\Omega = \{ -1, 0, 1, \ldots, 7\}$ from
    left to right along the $x$-axis. 
    The parameters used for this plot are $\ti{A} =
    0.1$, $\ti{\Gamma}_L = \ti{\Gamma}_R = 12$, $\ti{J}_L = \ti{J}_R =
    2.5$, $\tgext = 0.001$ and $\tTB = 3$. For negative $V$ and small
    positive $V$ we find an exponential decay corresponding to a
    thermal state. For larger $V > 0$ the distribution develops a
    peak at $n \not=0$ which indicates a driven state.}
  \label{fig:Pn}
\end{figure}

This is calculated in Fig.~\ref{fig:Pn} using the numerical approach
for different values of $eV/\hbar \Omega$. We find an exponential
decay for $V \leq 0$ corresponding to the high-temperature limit of
the Boltzmann distribution and a trend towards a driven state for $V >
0$.

In the regime where both resonances are red-detuned ($V < 0$, $-V <
V_G < V$), we find a cooling of the oscillator to temperatures well
below the bath temperature. This shows up in
Figs.~\ref{fig:StabilityV1} and \ref{fig:StabilityV3} as the little
triangular-shaped regions below the center, where the oscillator
energy drops below the value corresponding to the bath temperature.

Due to the non-linearity of the master equation, more than one
physical solution may emerge and we find that this is indeed the case
in the sector where the NR is strongly driven. An analogous effect was
found previously for the same system at the JQP cycle \cite{harvey08}
and for a more general class of
systems.\cite{flindt05,usmani07,isacsson04,novotny04,labarthe07,pistolesi04} We find that
generally, the response of the system close to a DJQP resonance is
much more pronounced than at the JQP in the sense that quantitatively
similar effects may be observed at much smaller values of the
coupling. This agrees with the prediction \cite{clerk05} that the
backaction effects at the DJQP exceed those of the JQP by a factor
$(\Gamma/J)^4$. 
Therefore the DJQP is favorable from the experimental
point of view since achieving a strong coupling is challenging.

\begin{figure}[t]
  \centering
  \includegraphics[width = 0.45 \textwidth]{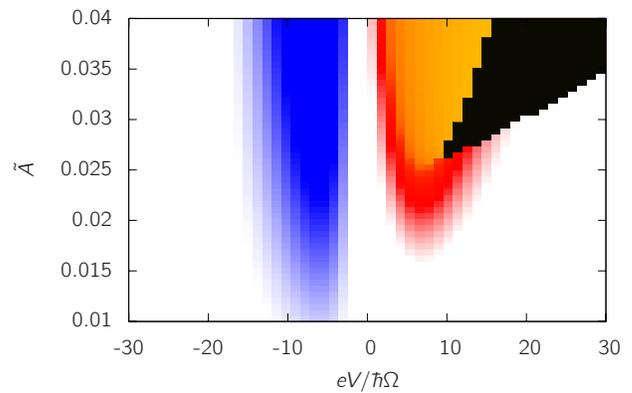}
  \caption{Oscillator energy in arbitrary units as a function of bias
    voltage $eV/\hbar\Omega$ and coupling $\ti{A}$ calculated in the
    Gaussian approximation. For $V <
    0$, backaction leads to cooling of the oscillator (blue). On the
    contrary, strong driving (red/yellow) can be observed for $V > 0$. Above
    a critical coupling $\ti{A}(V)$, the system enters a bistable
    region (black). The parameters are the same as in
    Fig.~\ref{fig:StabilityV1} and $eV_G/\hbar \Omega = 0$.}
  \label{fig:bistable}
\end{figure}

A plot of the location of the bistabilities found in the Gaussian 
approximation as a function of the bias
voltage $V$ and the coupling strength $\ti{A}$ is shown in
Fig.~\ref{fig:bistable}. Again, cooling of the oscillator is seen in
the blue regions for $V < 0$, whereas driving happens for $V >0$ as is
depicted by red regions. Towards stronger coupling, both effects
increase in magnitude. Two stable solutions appear only for a
blue-detuned SSET and the voltage range where such an effect is
visible grows with increased coupling. When increasing the coupling
for a given voltage $V > 0$ (which corresponds to a vertical cut in
Fig.~\ref{fig:bistable}), the system will evolve from a thermal state
via the bistable state to a single driven state. On the contrary, an
increase in voltage (corresponding to a horizontal cut) carries the
system from a thermal state to a driven state, then into the bistable
region. Beyond the bistable region, the system will fall back to the
thermal state. Note that the effect of driving is much stronger
than the cooling of the NR 
(cf.~Figs.~\ref{fig:StabilityV1} and \ref{fig:StabilityV3}).

We have confirmed the existence of bistability using the
numerical approach by explicitly calculating the complete probability
distribution $P(x)$ [results not shown explicitly]. 
In a thermal state, this distribution shows a
single peak at $x = 0$. In a driven state, on the contrary, two
symmetric peaks at finite values $|x| \neq 0$ appear. 
In the bistable regime, the system switches between these two states, 
leading to a distribution $P(x)$ that shows three peaks, 
one (thermal) at the origin and two side-peaks (driven) at $|x|>0$.
From these studies of the bistable regime using 
the numerical approach, we noticed that the parameter range 
in which the system exhibits bistability is smaller than 
the one obtained via the mean-field solution. 
The structure of the bistable region could be better predicted
using the analytical approach 
by including higher than second order cumulants, i.e.~extending
the analysis beyond the Gaussian approximation.

In the following sections, we will show that these different
states of the NR also manifest themselves in the transport properties
of the SSET, i.e.~the current and the current noise.

\section{Current properties}
\label{current}
We showed in the previous section that the coupling of an SSET to an
NR can drive the oscillator into a non-thermal state and effect in
cooling, potentially even cooling down close to the ground
state.\cite{clerk05, naik06} In the following, we will study if and
how it is possible to measure signatures of the resonator state in the
current and current noise characteristics of the SSET close to the
DJQP resonance.

The number of electrons that have left the island to the right lead,
$n_R$, is proportional to the transported charge and therefore
determines the current flow. Hence, the expectation value of the
current is given by 
\begin{align}\label{eq:current_nr}
  \expct{I}{} &= (-e) \frac{d}{dt} \expct{\hat n_R}{}
               = (- e) \Tr\{  \dot \rho(t)\; \hat n_R\}\;.
\end{align}
Without loss of generality, we chose to measure the current across the
\emph{right} junction. In the stationary limit, the total current is
conserved such that the currents across the left and right junctions
are equal. In each DJQP cycle, two tunneling events take place at the
right junction, see Fig.~\ref{fig:SSETDJQP}: the transfer of a
quasiparticle which takes the island from charge state $\ket{2}$ to
the state $\ket{1}$. Subsequently, a Cooper pair tunnels to the right
lead and leaves the island in the state $\ket{-1}$. Two processes
involving only changes in $n_L$ and $n$, which therefore do not
contribute to $\expct{I}{}$, close the cycle in which $3$ electrons in
total have been transported through the island.

Equivalently, in the stationary state the expectation value of the
current $\expct{I}{}$ can be written using the superoperator formalism. From Eq.~\eqref{drhototal}, one finds
\begin{align}
 \label{eq:current_Liouville}
 \expct{I}{} = (-e) \Tr_n \Tr_\osc \left( \mathcal{I}_{total} \rho_{stat} \right)
\end{align}
where $\mathcal{I}_{total} = \mathcal{I}_{qp} -2 \mathcal{I}_{CP}^+ + 2 \mathcal{I}_{CP}^-$ is the superoperator describing the total current.
Equation~\eqref{eq:current_Liouville} is used in this form in the numerical routine. 

For the analytic mean-field approximations we split
the total current into two terms, 
$\smash{\expcts{I}{} = \expcts{I^{ND}}{} + \expcts{I^{D}}{}}$, 
corresponding to 
a contribution from the tunneled Cooper pair (the non-dissipative
current) $I^{ND}$, and a contribution from the quasiparticle tunneling
event (the dissipative part) $I^D$. For these two contributions
we can write down the exact expressions
for the dissipative
\begin{subequations}\label{eq:current_D}
\begin{align}
 \expct{I^{D}}{} &=(-e) \Tr_n \Tr_\osc \left( \mathcal{I}_{qp} \rho_{stat} \right) \;, \\ &= (-e) \Gamma_R \expct{\hat p_{22}}{}\;, 
\end{align}
\end{subequations} 
and non-dissipative part
\begin{subequations}\label{eq:current_ND}
\begin{align}
\expct{I^{ND}}{} &= (-e) \Tr_n \Tr_\osc \left( [2 \mathcal{I}_{CP}^- -2 \mathcal{I}_{CP}^+] \rho_{stat} \right) \;, \\	&=(-e) 2 i J_R \left( \expct{\hat p_{1,-1}}{} - \expct{\hat p_{-1,1}}{} \right)\;.
\end{align}\end{subequations}

In the thermal-oscillator approximation, these expectation
values can be calculated by solving for the corresponding elements of
the density matrix, as shown in detail in the
Appendix~\ref{app:thosc}. We find that the vector of all finite 
$\expct{\hat p_{kj}}{}$, $\ket{\mathbf p}$, is given by $\ket{\mathbf
  p} = i \ti{J}_L \mathbf{M}^{-1} \ket{c}$ where $\mathbf{M}$ is the evolution matrix of the SSET system containing all
the system parameters, Eq.~\eqref{eq:M}, and the constant $\ket{c}$ is
the inhomogeneous part of the master equation due to the
normalization of the density matrix $\sum_k \expct{\hat p_{kk}}{} = 1$.
Using this result the stationary current for the DJQP cycle can be
written as
\begin{align}
  \expct{I}{} &=  \frac{3}{2} (-e) \Omega \left[ 
                    \frac{1}{\ti{\Gamma_R}} + \frac{1}{\ti{\Gamma_L}} 
                  + \frac{1}{\gamma_L(\tx)} + \frac{1}{\gamma_R(\tx)}
                 \right]^{-1}
\label{eq:current_rate}\;.
\end{align}
The inverse of the tunneling rates for quasiparticles and Cooper
pairs, $\Gamma_{L,R}$ and $\gamma_{L,R}$, respectively, can be
interpreted as effective resistances for these processes. Then,
Eq.~(\ref{eq:current_rate}) is reminiscent of the current through a
series of resistors, where the largest resistance determines the
behavior. The Cooper pair tunneling rates are given by \cite{averin89}
\begin{align}\label{defCooperRates1}
  \gamma_L(\tx) &= 2 \ti{\Gamma_R} \frac{\ti{J_L}^2}
                                   {(\ti{\Gamma_R}/2)^2 + \epsilon_{2,0}^2(\tx)}
\;, \\
  \gamma_R(\tx) &= 2 \ti{\Gamma_L} \frac{\ti{J_R}^2}
                                   {(\ti{\Gamma_L}/2)^2 + \epsilon_{1,-1}^2(\tx)}\;. \label{defCooperRates2}
\end{align}
where $\epsilon_{k,j}$ denotes the difference in energy between the charge states $\ket{k}$ and $\ket{j}$ and thus measures the detuning from the DJQP resonance. The renormalized tunneling rates of the SSET are defined by
$\ti{\Gamma}_\alpha = \Gamma_\alpha/\Omega$ and $\ti{J}_\alpha =
J_{\alpha}/2 \hbar \Omega$. If the Cooper pair tunneling, say, to the
right lead is resonant, i.e.~$\epsilon_{1, -1} = 0$, the rate
$\gamma_R$ reaches a maximum at the value $\gamma_R = 8
\ti{J_R}^2/\ti{\Gamma_L} = (2 J_R^2/\Gamma_L)/\Omega$. It decays like
a Lorentzian away from the resonance. Expressions for the current in
less general form are for example derived for $\epsilon_{jk} = 0$ in
Ref.~[\onlinecite{clerk03}] and for $|\epsilon_{1,-1}| =
|\epsilon_{2,0}|$ in Ref.~[\onlinecite{clerk02}].

Due to the capacitive coupling of the SSET to the NR, the resonance is
shifted compared to the uncoupled case in the thermal
approximation. We find in dimensionless units (derivation given in
Appendix~\ref{app:thosc})
\begin{align}
  \epsilon_{1,-1}(\tx) &= \frac{e V_G}{\hbar \Omega}  
                       + \frac{e V}{\hbar \Omega}
                       - 2 \ti{A} \expct{\tx}{}\;, 
\label{eq:def_e1m1} \\  
  \epsilon_{2,0}(\tx) &= \frac{e V_G}{\hbar \Omega}  
                       - \frac{e V}{\hbar \Omega}
                       - 2 \ti{A} \expct{\tx}{}\;.
\label{eq:def_e20}
\end{align}
where $eV/\hbar \Omega$ and $e V_G/\hbar \Omega$ are
the relative bias and gate voltages measured from 
the values at the DJQP resonance.

The SSET is affected only if the average position of the NR is finite,
i.e.~$\expcts{\tx}{} \not= 0$. This shift in the equilibrium position of the NR effectively corresponds, from the point of view of the SSET, to a change in
$V_G$ and will therefore be referred to as an \textit{effective backgate behavior} later on. This effect is of second order in the coupling $A$ since we
observed in the previous section that $\expcts{\tx}{}$ is linear in $A$.

Note that the average displacement of the NR oscillation 
in the stationary limit, 
\begin{align}\label{eq:x_thermal}
  \expct{\tx}{} &= 
  2 \ti{A} \expct{n}{} = 2 \ti{A}
                \left( \frac{1}{\ti{\Gamma}_R} + \frac{1}{\gamma_R(\tx)}\right)
                \frac{2\expct{I}{}}{3(-e)\Omega}\;,
\end{align}
is determined by the Cooper pair tunneling rate $\gamma_R$ in addition
to the quasiparticle tunneling rate $\ti{\Gamma}_R$. This is in
contrast to the JQP cycle where it is only the necessarily small
$1/\ti{\Gamma}_R$ which determines the displacement. Since the rates
and the current are implicitly dependent on $\expcts{\tx}{}$ via the
Cooper pair tunneling rate, Eq.~\eqref{eq:x_thermal} is a
self-consistency equation.

\begin{figure}[t]
  \centering
  \includegraphics[width = 0.45 \textwidth]{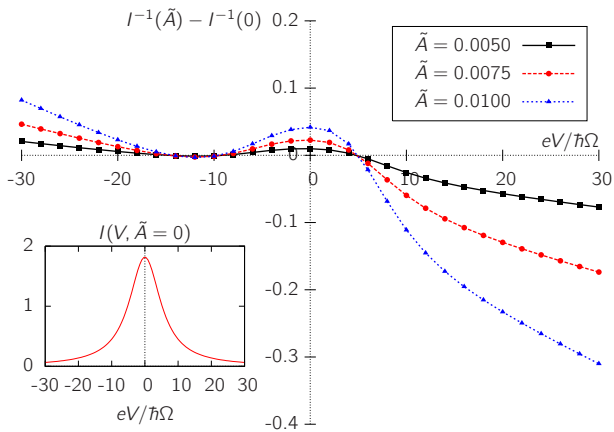}
  \caption{Difference of the inverse current 
$1/\expcts{I}{}(\ti{A}) - 1/\expcts{I}{}(\ti{A}=0)$ versus bias voltage 
$eV/\hbar \Omega$ for various coupling
strengths $\ti{A}$. From the intersection with the voltage axis, the 
shift $2 \ti{A} \expct{\tx}{}$ 
of the oscillator position can be read out. Lines
correspond to the numerical analysis and agree well with the 
Gaussian approximation (points) in this parameter regime.
Inset: Average current $\expct{I}{}$ through the SSET for the parameters
$\ti{\Gamma}_L=\ti{\Gamma_R}=10$, $\ti{J}_L = \ti{J}_R = 2$, 
$\tgext = 10^{-4}$, and $\tTB = 3$. }
  \label{fig:VerenaPlot1}
\end{figure}

Going beyond the thermal-oscillator approximation, we use
again the truncated master equation and the numerical approach to
calculate the current via the general Eqs.~\eqref{eq:current_D} and
\eqref{eq:current_ND}. In order to assess the quality of the
Gaussian approximation, we first compare the results of the
two approaches for low coupling strength. The result is shown in
Fig.~\ref{fig:VerenaPlot1}, where we plot the difference in the
inverse current between the weakly coupled and the uncoupled
system. The results of the Gaussian approximation and the
numerically evaluated lines are in excellent agreement.
The Lorentzian lineshape of the current (inset of Fig.~\ref{fig:VerenaPlot1})
is preserved in case of the weak coupling.

The change in the average current due to the coupling to the oscillator can be most transparently illustrated by plotting the
difference of the inverse currents in the coupled and the uncoupled cases, see Fig.~\ref{fig:VerenaPlot1}. As obvious from Eq.~\eqref{eq:current_rate}
the inverse current $1/\expct{I}{}$ is given by the sum of rates involving
the various transport processes. 
In the thermal oscillator approximation, the function
$1/\expct{I}{}(\ti{A}) - 1/\expct{I}{}(\ti{A} = 0)$ 
changes sign as a function of $eV/\hbar\Omega$ at a position which is proportional to $2 \ti{A} \expcts{\tx}{}$ as is shown in Eq.~\eqref{eq:app_1overcurrent}.
We expect this sign change to be the most feasible way 
to experimentally observe the influence of the NR on the SSET current and to investigate quantitatively the coupling strength using only the average current.

\begin{figure}[t]
  \centering
  \includegraphics[width = 0.45 \textwidth]{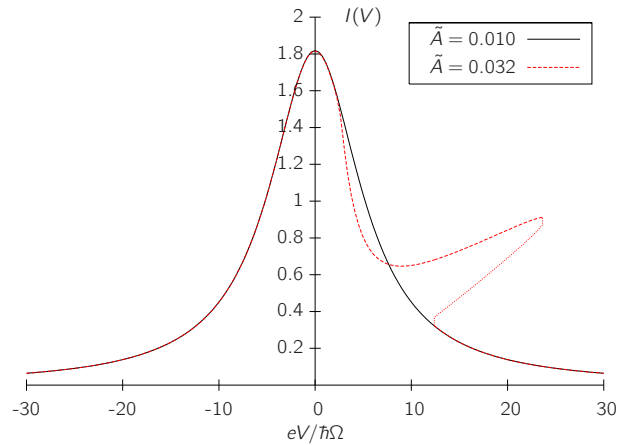}
  \caption{Average current $\expct{I}{}$
    versus bias voltage for $\ti{A} = 0.01$ and
    $\ti{A} = 0.032$, $\tTB = 2.5$, further parameters as in Fig.~\ref{fig:VerenaPlot1}. For increased coupling the Lorentzian
    peak becomes distorted and two stable solutions emerge which
    result in two stable values for the current.}
  \label{fig:SShapedCurrent}
\end{figure}

The bistability of the oscillator states, which was already discussed
in the previous section, also manifests itself in the current through
the SSET.  For increased coupling, we find that the equation of motion derived within the Gaussian approximation has up to
three solutions, of which two correspond to stable currents. The
resulting current-voltage characteristic is shown in
Fig.~\ref{fig:SShapedCurrent}. In the center, the usual DJQP resonance
is clearly visible. While for very low coupling $\tA \approx 0.01$,
the current still follows approximately a Lorentzian, strong
deviations become visible already for $\tA=0.032$.

In an experimental setup, we do not expect two stable currents to be
distinguishable. Indeed, a current measurement of the SSET-resonator system
will yield a weighted average\cite{flindt05b} because decoherence effects will lead to a
switching between the two stable configurations on time scales large
compared to the oscillation period but small compared to the
measurement resolution.\cite{jordan04} These switching rates can
easily be inferred from a comparison of the measured current to the
two stable values and from the current noise, as we shall show 
in section \ref{noise}.

\section{Noise properties}
\label{noise}

\subsection{Charge noise}
In the past, linear-response arguments \cite{clerk05} have been used to
support the idea that a generic detector acts on the resonator in the
same way as a second thermal bath, and that the backaction on the
resonator caused by charge fluctuations on the island can be
described essentially by two parameters, a damping rate $\geff$ and an
effective temperature $k_B \Teff$. These are related to the charge-
fluctuation spectrum
\begin{align}
   S_n(\omega) = \int_{-\infty}^\infty dt\; e^{i \omega t}
\cumu{ n(t) n(0) }{}\;.
\end{align}
via
\begin{align}\label{eq:geff}
   \ti{\gamma}_\mathrm{eff}(\omega) &= \ti{A}^2
        \frac{S_n(\omega) - S_n(-\omega)}{2 \omega} \;, \\
\label{eq:Teff}
   \ti{T}_\mathrm{eff}(\omega) &= \ti{A}^2 \frac{S_n(\omega) + S_n(- 
\omega)}{2 \ti{\gamma}_\mathrm{eff}(\omega)}\;,
\end{align}
in the limit $k_B \Teff(\omega) \gg \hbar \omega$, where $\omega$ is  
given
in units of $\Omega$, $\ti{T}_\mathrm{eff} = k_B \Teff/\hbar \Omega$
and $\ti{\gamma}_\mathrm{eff} = \geff/\Omega$.
Since these expressions follow from a linear-response calculation, both
effective quantities are written in terms of the bare charge-noise,
calculated in the absence of coupling with the oscillator.

Investigating the retarded and advanced (absorption and emission)
contribution of the charge correlation explicitly, we can derive an
analytic expressions for $S_n(\omega)$ for the
uncoupled SSET (see Appendix~\ref{app:charge})
\begin{align}
   \frac{S_n(\omega) - S_n(- \omega)}{2 \omega} &=
2 i \langle n | \big[ \omega^2 + \mathbf{M}^2 \big]^{-1} \mathbf{K}_- |
\mathbf{p} \rangle\;,
\label{eq:Sn_asym}
\\
   \frac{S_n(\omega) + S_n(- \omega)}{2} &=
2 \langle n | \mathbf{M} \big[ \omega^2 + \mathbf{M}^2 \big]^{-1}
\left( \mathbf{K}_+ - \expct{n}{}\right) | \mathbf{p} \rangle\;,
\label{eq:Sn_sym}
\end{align}
where $\mathbf{M}$ denotes again the evolution matrix of the SSET
system, $\bra{n}$ is defined such that $\langle n | \mathbf p \rangle
= \expct{n}{}$ and $\mathbf{K}_\pm$ denote coupling matrices given
explicitly in Eqs.~\eqref{eq:Kplus} and \eqref{eq:Kminus}. Note that
$\cumu{n^2}{} = \expcts{n^2}{} - \expcts{n}{}^2= \bra{n} \mathbf{K}_+
\ket{\mathbf p} - \expcts{n}{}^2$ and $\mathbf{K}_-$ acts only on the
off-diagonal elements of $\expcts{\hat p_{kj}}{}$, i.e.~the Cooper pair
tunneling terms.

The self-consistency equation for 
$\cumu{\tx^2}{}$, that has to be solved in the Gaussian approximation,
can be written as
\begin{align}
   \cumu{\tx^2}{} &=
\frac{2 \gext T_B + 2 \gSSET \TSSET(\tx, \tx^2)}
      {\gext + \gSSET(\tx, \tx^2)}\;,
\label{eq:x2_Gaussian}
\end{align}
where
\begin{align}
   &\gSSET(\tx, \tx^2) = i (2 \tilde{A})^2 \\ & \quad
     \langle n | \big(\mathbf{1} - \tgext (\tgext + \mathbf{M}) \big)
        \big( \mathbf{1} + \tgext \mathbf{M} + \mathbf{M}^2 \big)^{-1}
        \mathbf{K}_- | \mathbf{p} \rangle\;,
\label{eq:gSSET} \notag \\
&  2 \gSSET \TSSET(\tx, \tx^2) = (2 \tilde{A})^2 \\ & \quad
     \langle n |  \big( \tgext + \mathbf{M} \big)
        \big( \mathbf{1} + \tgext \mathbf{M} + \mathbf{M}^2 \big)^{-1}
        \big( \mathbf{K}_+ - \langle n \rangle \big) | \mathbf{p}  
\rangle \notag ;
\label{eq:TSSET}
\end{align}
We observe that the mean-field equation in second order provides the
same physics as linear-response theory,
i.e.~$\ti{\gamma}_\mathrm{eff}$ at $\omega/\Omega = 1$ is 
of the same form  as $\gamma_\mathrm{SSET}$. 
Since Eq.~\eqref{eq:x2_Gaussian} is a self-consistency equation for 
$\cumu{\tx^2}{}$ and not for the effective oscillator energy as in 
Eqs.~\eqref{eq:geff} and \eqref{eq:Teff} the expressions differ by a
factor of $4$. The result in Eq.~\eqref{eq:x2_Gaussian} is more
accurate in the sense that the parameters of the damped oscillator
are involved: $\mathbf{1} + \tgext \mathbf{M} + \mathbf{M}^2 = \big(
\mathbf{1} - (\tgext/2)^2 \big) + \big( \mathbf{M} + \tgext/2 )^2$
with a renormalized frequency of $\Omega_r = \sqrt{\Omega^2 -
  (\gext/2)^2}$ and additional damping due to $\gext/2$.

Note that Eq.~\eqref{eq:x2_Gaussian}
is a self-consistency equation for $\expcts{\tx^2}{}$
since $\ket{\mathbf{p}} = |\mathbf{p}(\tx, \tx^2) \rangle$
and it has to be solved together with $\expcts{\tx}{} = 2 \ti{A}
\expcts{n}{}$. Even if it is assumed that $\ket{\mathbf p} = |
\mathbf{p}(0, 2 \tTB) \rangle$, the expression contains a correction
due to the finite quality factor of the NR.

Whereas the approach describing the detector as an effective bath
proved very successful in providing a simple physical explanation of
experiments,\cite{naik06} some of its shortcomings have started to be
identified in recent theoretical works.\cite{bennett08,rodrigues08}
For example, it has very recently been proposed
\cite{rodrigues08} that the signature of the oscillator in the charge
noise spectrum of a generic detector is \emph{not} the one of a
thermal oscillator. In the light of these findings, the calculation
of the full frequency-dependent charge noise spectrum of the SSET near
the DJQP in the presence of an oscillator becomes relevant, even more
so since the charge-noise spectrum is an experimentally accessible
quantity. 

As shown in Appendix~\ref{app:charge}, it is possible to use the
master-equation approach to derive formal expressions for
$S_n(\omega)$, at least for weak coupling in the thermal-oscillator
approximation. However, it turns out that these expressions are
difficult to evaluate explicitly for stronger couplings (in the
Gaussian approximation). On the other hand, the fully numerical
approach presented in section~\ref{liouville} can easily be adapted to
allow the calculation of finite-frequency correlation functions of
\textit{system} (as opposed to \textit{bath}) operators, using the
quantum regression theorem.\cite{gardinerzoller,flindt05b} In the
following, we therefore discuss only the charge-noise spectrum
obtained numerically. Note that we verified that our algorithm
reproduces accurately the known charge and position fluctuation
spectra in the uncoupled ($\tilde{A}=0$) regime.

\begin{figure}[t]
   \centering
   \includegraphics[width=0.45 \textwidth]{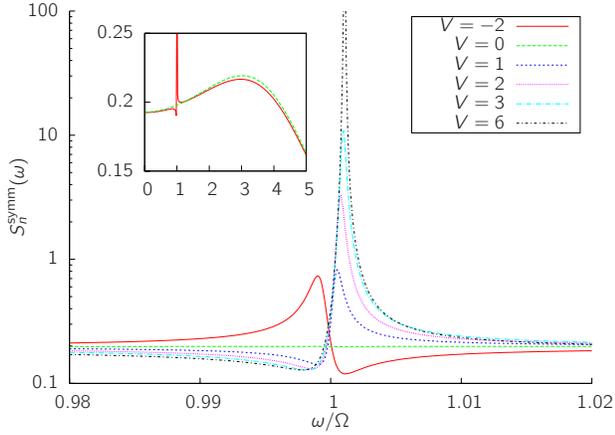}
\caption{Frequency-dependent charge noise of the charge $n$ on the
  SSET, symmetrized contribution $S^\mathrm{symm}_n(\omega)$, versus
  $\omega/\Omega$ for different values of the bias voltage $eV/\hbar
  \Omega$ measured from the resonance. 
  Parameters are identical to Fig.~\ref{fig:Pn}. While there
  is no structure in the case $V = 0$, we can distinguish if the
  oscillator is driven or cooled by the symmetry of the peak at
  $\omega = \Omega$. Inset: $S_n^\mathrm{symm}(\omega)$ in a larger parameter
  regime; numerical calculation $\ti{A} = 0.1$ compared to analytical
  result $\ti{A} = 0$.}
  \label{fig:Snw}
\end{figure}

Figure~\ref{fig:Snw} shows the symmetrized (in frequency) 
charge-noise spectra, 
$S^\mathrm{symm}_n(\omega) = [S_n(\omega) + S_n(-\omega)]/2$, 
obtained for different values of the bias-voltage
detuning from the DJQP resonance. The oscillator state, i.e.~thermal or
driven, can be determined from the Fock space probability distribution in
Fig.~\ref{fig:Pn}.  The inset shows that the signature of the
oscillator in $S^ \mathrm{symm}_n(\omega)$ appears prominently around
the natural frequency of the oscillator. Away from $\omega\sim
\Omega$, the charge spectrum is only weakly affected by the
oscillator, since the coupling of the island to the resonator changes
the effective biasing conditions of the SSET.

The main panel of Fig.~\ref{fig:Snw} shows the evolution of the
charge-noise spectra when the system is taken from the ``cooling''
region ($V=-1$) through the resonance point ($V=0$), to the voltage
regime where the state of the oscillator becomes highly non-thermal.
Unsurprisingly, the overall signal around $\Omega$ increases
dramatically when the oscillator enters the driven regime, reflecting
the overall increase in the magnitude of $S_x (\omega\sim \Omega)$
when the oscillators energy is increased. Associated with this
increased magnitude is an overall reduction of the linewidth, that is
again explained rather straightforwardly via the decreased total
damping rate in this region ($\gamma_{\mathrm{eff}}<0$).

The most interesting observation to be made about
Fig.~\ref{fig:Snw} is perhaps the striking similarity between the
spectra at $V<0$ and $V\gg0$. In both cases, we find not only a
resonance at the renormalized frequency of the oscillator, but also a
dip at its bare frequency, exactly as derived in
Ref.~[\onlinecite{rodrigues08}] for a generic detector. Note that the
renormalized frequency 
$\Omega_r(V) = \sqrt{\Omega^2 - [(\gext + \geff(V))/2]^2}$ 
depends on the detuning $V$.  For $V < 0$ we
find $\gamma_{\mathrm{eff}}>0$, such that the renormalized frequency
$\Omega_r$ is smaller than $\Omega$. 
In the region $V > 0$ and $\gamma_\mathrm{eff} < 0$,
the situation is reversed and the resonance appears at frequencies
$\Omega_r$ higher than $\Omega$, while the dip is pinned.
No structure is observed exactly at $V=0$ 
which relates to $\gamma_{\mathrm{eff}}(V=0) = 0$ since
absorption and emission of energy from the SSET to the oscillator
at the DJQP resonance is equal.

The presence of a dip is not compatible with a purely
``thermal'' state of the oscillator, even in cases where the Fock state probability distribution function $P(n_\mathrm{osc})$ decays exponentially
like in the fully-thermal case [cf.~Fig.~\ref{fig:Pn}]. Not only does
this result confirm that the simple model used in
Ref.~[\onlinecite{rodrigues08}] also applies to the complex
SSET-resonator system, it also demonstrates that the ``Fano-like''
lineshape, where both a resonance and a dip appear in the
spectrum of the charge noise, characterizes nicely the charge noise
spectrum on both the ``driving'' and ``cooling'' sides of the
resonance.

\subsection{Current noise}

We argued previously that the non-thermal oscillator states (i.e.,
cooling and driving of the resonator) manifest themselves in the
cumulants of coupled system. Therefore, it is reasonable to expect that
the coupling with the NR will modify the current fluctuations of the SSET.

In the following, we will focus mainly on the Fano factor, i.e.~the current
noise at $\omega = 0$ which is easily accessible by standard noise
measurement techniques. The
frequency-dependent current noise is given by MacDonald's
formula \cite{macdonald49}
\begin{align}
  S_I(\omega) 
&= \int dt e^{i \omega t} \cumu{
	\{ I(t), I(0) \} }{} \notag \\
&= (-e)^2 \omega \int\limits_0^\infty dt \sin(\omega t)
                               \frac{d}{dt} \cumu{n_R^2(t)}{}\;.
\end{align}
From this formula, we can deduce the following zero-frequency limit,\cite{rodrigues05b, flindt05b}
\begin{align}
  S_I(\omega \to 0) &= (-e)^2 \lim_{t \to \infty} \frac{d}{dt}  
\cumu{n_R^2(t)}{} \;.
\end{align}
Thus we have to determine the long-time limit of $\tfrac{d}{dt}
\cumu{n_R^2(t)}{}$. Note that we
assume for the derivation symmetric capacitances such that $S_{LL}(\omega = 0) =
S_{RR}(\omega = 0)$ due to charge conservation.\cite{armour04}
Since we use the symmetrized current noise, the Fano factor is defined
as $F = S_I(\omega = 0)/ (-e) \expct{I}{}$ without a factor of $2$.

In order to use Eq.~(\ref{drhototal}) to calculate numerically the current noise, we follow closely the approach presented in Ref.~[\onlinecite{flindt05}]. The original approach which applied to incoherent processes can be generalized to the coherent Josephson tunneling. The zero-frequency current noise is thus 
given by
\begin{align}
 S_I(\omega \rightarrow 0) 
&= (-e)^2 \Big[ \Tr_n \Tr_\osc (\mathcal{I}_{noise} \rho_{stat}) \notag \\
&- 2 \Tr_n \Tr_\osc (\mathcal{I}_{total} \mathcal{R} \mathcal{I}_{total} \rho_{stat}) \Big]
\label{eq:noise_L}
\end{align} 
where $\mathcal{I}_{total} = \mathcal{I}_{qp} -2 \mathcal{I}_{CP}^+ + 2 \mathcal{I}_{CP}^-$ is the superoperator describing the total current as defined previously, $\mathcal{I}_{noise} = \mathcal{I}_{qp} + 4 \mathcal{I}_{CP}^+ + 4 \mathcal{I}_{CP}^-$ and $\mathcal{R}$ is the pseudoinverse of the Liouvillian $\mathcal{L}$.
\cite{flindt05b}
We compared both the Fano factor and the frequency-dependent
current noise in the uncoupled case with the exact expressions of 
Ref.~\onlinecite{clerk03} (see also App.~\ref{app:noise})
and verified thus the correctness of Eq.~\eqref{eq:noise_L}.

As in the previous section it is possible to obtain an analytic
expression in the thermal-oscillator approximation. For
details of the calculation we refer to Appendix~\ref{app:noise}. We
find
\begin{align}
 F &= \frac{3}{2}  + 
\frac{3}{2}\frac{-f_\mathrm{sym} + f_\mathrm{asym}}
{\left(\dfrac{1}{\ti{\Gamma_L}} + \dfrac{1}{\ti{\Gamma_R}}
    + \dfrac{1}{\gamma_L} + \dfrac{1}{\gamma_R}
\right)^2}
\ .
\end{align}
The term
\begin{align}
f_\mathrm{sym} =3 \left(\dfrac{1}{\ti{\Gamma_L}}  
+ \dfrac{1}{\ti{\Gamma_R}}  
\right)
\left( \dfrac{1}{\gamma_L}  + \dfrac{1}{\gamma_R} \right)
\end{align}
reduces the noise \cite{choi01,choi03} at the resonance; 
this is a consequence of the coherence of the Cooper pair tunneling.
The asymmetric part 
\begin{align}
f_\mathrm{asym} &= \left( \frac{1}{\ti{\Gamma_L}} - \frac{1}{\ti{\Gamma_R}}  
\right)^2
       + \left( \frac{1}{\ti{\Gamma_L}} - \frac{1}{\ti{\Gamma_R}}  
\right)
         \left( \frac{1}{\gamma_L} - \frac{1}{\gamma_R} \right) \notag
\notag\\ &+ \left( \frac{1}{\gamma_L} - \frac{1}{\gamma_R} \right)^2
+ \frac{\epsilon_{2,0}^2}{(\ti{\Gamma_R}/2)^2 \ti{J_L}^2}
+ \frac{\epsilon_{1,-1}^2}{(\ti{\Gamma_L}/2)^2 \ti{J_R}^2}
\end{align}
increases the noise level, for example if the coupling
is asymmetric, $\ti{\Gamma}_L \not= \ti{\Gamma}_R$, 
or the system
is tuned away from the resonance, i.e.~ $\epsilon_{2,0} \not= 0$ or 
$\epsilon_{1,-1} \not= 0$.

\begin{figure}[t]
  \centering
  \includegraphics[width = 0.45 \textwidth]{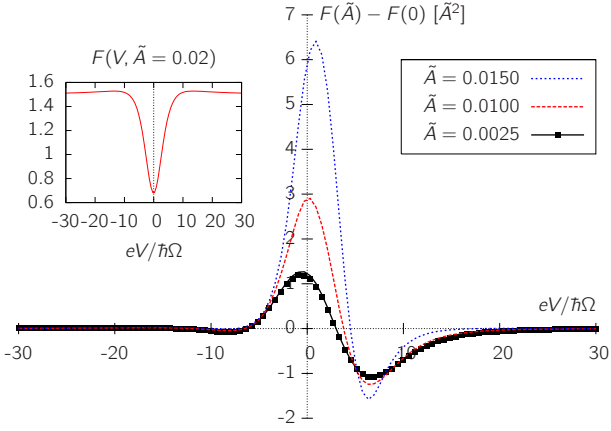}
  \caption{Difference in Fano factors $\Delta F = F(\ti{A}) - F(\ti{A}=0)$ 
in units of $\ti{A}^2$ as a function of bias voltage
$eV/\hbar \Omega$ for $\ti{\Gamma}_L =\ti{\Gamma}_R = 10$,
$\ti{J}_L = \ti{J}_R = 2$, $\tgext = 0.0001$ and $\tTB = 2.5$. Lines depict results
of the numerical approach. The results of the Gaussian approximation (points) fit
quantitatively only for $\ti{A} < 0.01$. Inset: the Fano factor $F(\ti{A} = 0.02)$ decreases at $V \approx 0$ when the Cooper pair tunneling is resonant.} \label{fig:Fano1}
\end{figure}

The coupling of the SSET and the NR enters via the $\expcts{x}{}$
dependence of the Cooper pair tunneling rates $\gamma_L(x)$ and
$\gamma_R(x)$ defined in Eqs.~\eqref{defCooperRates1} 
and \eqref{defCooperRates2}. In the
thermal-oscillator approximation the effect is thus of
second order in $A$ as observed already for the current. In analogy to
the previous section the coupling of the SSET to the NR leads in the
thermal approximation only to a shift in the resonance position of the
SSET. Since the resonance condition is important for coherent
Cooper pair tunneling, the shift manifests itself in a higher
noise signal around $V = 0$, leading to a peak in $\Delta F = F(\ti{A} \not= 0) - F(\ti{A} = 0)$, see Fig.~\ref{fig:Fano1}. At large voltages both the Fano factor of 
the coupled and uncoupled SSET converge to the value $3/2$.

In order to access the regime of increased coupling, we investigated
the current noise also in the Gaussian approximation and compared the result to
the numerical calculation explained below. For very low
coupling up to $\ti{A} = 0.01$, as long as the resonator remains close to
its thermal state, the two approximations coincide. For increased
coupling, however, deviations from the \textit{effective backgate} behavior 
appear. A
plot of the difference in Fano factors $\Delta F$ as a function of the
bias voltage is shown in Fig.~\ref{fig:Fano1}.

We find that the central peak in $\Delta F$ is accompanied by two
negative side peaks where the current noise of the coupled SSET is
lower than the uncoupled value. In the cooling and driving regimes the
SSET absorbs and emits energy to the resonator in order to move
closer to the resonance position. As pointed out before, the noise at
the resonance is reduced due to the coherent Cooper pair tunneling and
therefore the cooling and driving mechanisms are responsible for the
negative side peaks. As in the energy of the resonator, Fig.~\ref{fig:bistable},
the driving is stronger than the cooling, which
manifests itself in the heights of the negative side peaks.

\begin{figure}[t]
  \centering
  \includegraphics[width = 0.45 \textwidth]{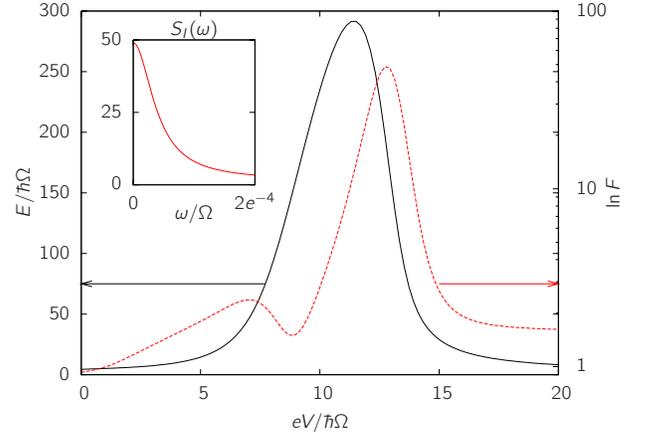}
  \caption{Logarithmic Fano factor $\ln F$ and oscillator 
        energy $E/\hbar \Omega$ as a function of bias voltage $eV/\hbar \Omega$.
	The parameters are $eV_G/\hbar \Omega = 2.5, \ti{\Gamma}_L =\ti{\Gamma}_R = 12$,
$\ti{J}_L = \ti{J}_R = 2.5$, $\tgext = 0.001$, $\tTB = 3$, and $\tA = 0.1$. 
The Fano factors
shows a first peak when the system enters the driven state. At the onset of
the bistability, the Fano factor increases drastically on the logarithmic scale,
 which we attribute
to telegraph noise due to switching between the two stable configurations.
Inset: frequency-dependent current noise in the bistable region ($eV/\hbar\Omega = 12.8$) which decays on the scale of the switching rates.}
  \label{fig:FanoBistable}
\end{figure}



Even before the bistability arises, the coupling between the SSET and the NR leads to a modification of the
current noise, from which one can deduce if (and to which extent) the oscillator is cooled or driven.

In the bistable regime, the master equation approach leads to two
solutions for the noise corresponding to the two stable solutions for
the current. We argued previously that in this regime, a thermal and a
driven solution coexist.
Experimentally, however, the two solutions will not be completely
stable and the system will switch between the two states 
on a time scale much slower than the other system time scales. As was shown
generally for bistable systems,\cite{flindt05,usmani07} and was also
recovered for the case of the JQP,\cite{harvey08} the experimentally
measured noise will then be dominated by the switching between the two
stable states and will then essentially be telegraph noise. This type
of low-frequency noise is most clearly visible in the Fano factor which we
have evaluated using the numerical approach in Fig.~\ref{fig:FanoBistable}.

The Fano factor first peaks roughly at the transition between the thermal
state and the driven state. It then descends to a local 
minimum. This behavior is correlated to the stationary value of the
average position of the oscillator, $\expct{x}{}$, which acts as 
an effective backgate on the SSET as discussed in detail
in connection with Fig.~\ref{fig:Fano1}. In the driven state the transfer of
energy from the SSET to the NR is maximal and therefore we observe
a minimum in the Fano factor. Beyond the pure driven state,
the system reaches the bistable region which clearly shows up as a 
drastically increased Fano factor on the logarithmic scale 
in Fig.~\ref{fig:FanoBistable}.  
We attribute this zero-frequency noise feature
to a slow switching between the two stable configurations of the system.\cite{harvey08}

The inset of Fig.~\ref{fig:FanoBistable} shows the frequency-dependence
of the current noise 
at the voltage where the Fano factor is maximal. As expected
for a bistable system the value drops fast from the superpoissonian value on
a scale given by the sum of the two switching rates~\cite{flindt05b}, which
are very slow compared to the other scales of the system. In the 
frequency-dependent current noise [not explicitly shown here] 
we find signatures at the frequency of the oscillator
and at higher harmonics, which appear as a thermal resonance on
top of the uncoupled current noise of the SSET.


\section{Conclusions}
\label{conclusion}
In conclusion, we have provided a comprehensive treatment of a
superconducting single-electron transistor capacitively coupled to a
nanomechanical resonator. Assuming a linearized coupling for typical
system parameters, we have found that signatures of the mutual
interaction manifest themselves as well in physical quantities related
to the oscillator as in the current and noise properties of the
transistor.

It is known that this setup with the SSET close to the resonance
condition for the double Josephson quasiparticle process is especially
suitable to cool the oscillator or drive it. We confirm this
behavior and explain how the different approximations fail or succeed
in capturing these features. We use two complementary
approaches, a purely numerical solution of the Liouville equation
and a mean-field analysis, which lead to identical results in the
accessible parameter region. In addition
to driving effects of the oscillator, our solution predicts the
emergence of a bistable regime for increased coupling.

Although the consequences of cooling and driving the resonator have
been studied in previous work, the questions of 
if and how these non-thermal states can be measured in the
transport properties of the coupled SSET close to the DJQP resonance
were not discussed in the literature. This obvious gap is 
addressed in our work. We find that the current through the
SSET gives a measure of the displacement of the oscillator since, at the
lowest-order, the coupling with the NR modifies the gating condition of the SSET.

Investigating the fluctuations close to the 
resonance frequency of the mechanical oscillator, we find
pronounced effects. For instance,
the frequency-resolved \textit{charge} noise shows a sharp
resonance/dip structure close to the resonator frequency,
which allows to estimate the sign of the effective damping. Moreover,
we have studied the zero-frequency \textit{current} noise, i.e.~the Fano factor.
Depending on the applied voltages, the SSET coupled to an NR results in an 
increase or decrease of the Fano factor compared
to the uncoupled case. We could show that a decrease
is related to driving or cooling of the NR. Furthermore we find
that the Fano factor increases notably as soon as the system enters the bistable
state. This feature can be used experimentally to pinpoint the bistable
region. The switching rates between the two stable regimes 
can be identified from the frequency dependence
of the current noise.
 
In comparison to an SSET driven at a JQP resonance, the DJQP shows
stronger effects for lower couplings. Extending the parameter regime
to strong coupling, different methods have to be applied and also the
assumption of a linear coupling is challenged. This will be the
subject of future work.

\acknowledgments 
We would like to thank A.~D.~Armour and C.~Flindt
for interesting discussions and M.~S.~Choi for his support. This
work was financially supported by the Swiss NSF and the
NCCR Nanoscience.

\appendix

\section{Coupling of SSET and NR}
\label{app:coup}

The coupling between the SSET and the NR originates from a capacitive
coupling between the island with the gated resonator. The displacement
$x$, i.e.~the deviation from the equilibrium distance $d$, determines
the capacitance $C_N(x)$
\begin{align}
  C_N(x) = \epsilon_0 \frac{A}{x + d} 
         = C_N^0 - C_N^0 \frac{x/d}{1 + x/d} \;,   
\end{align}
and thus the charging energy $E_C$ and the charge on the island $n_0$
\begin{align}
E_C(x) &= \frac{e^2}{2} \frac{1}{C_L + C_R + C_G + C_N(x)} \notag \\
       &= E_C^0 \frac{1}{1 - \frac{C_N^0}{C_\Sigma} \frac{x/d}{1 + x/d}} \;,\\
n_0(x) &= \frac{1}{e} \big( C_L V_L + C_R V_R + C_G V_G + C_N(x) V_N\big) 
\notag \\
       &= n_0^0 - \frac{C_N^0 V_N}{e} \frac{x/d}{1 + x/d} \;.
\end{align}
The linear expansion of the charging term of the Hamiltonian $H_C$ in
$x$ yields
\begin{align}\label{eq:Ecx}
  E_C(x) \big[ n + n_0(x) \big]^2 
  &=  E_C^0 \big( n + n^0_0 \big)^2 \\
   &+ 2 E_C^0  \big( n + n^0_0 \big) \frac{C_N^0 V_N}{e} 
     \Big(- \frac{x}{d}\Big)\;.\notag
\end{align}
We neglect a shift of the charging energy of the order of $E_C
C^0_N/C_\Sigma$ since in typical experimental setups,\cite{naik06}
this prefactor is much smaller than the term in the second line 
of Eq.~\eqref{eq:Ecx}. This
term would be proportional to $x (n + n_0)^2$ and would describe a
slightly different coupling.

Since $n^0_0$ constitutes a constant energy offset it is
neglected in the calculation. The interaction constant in $H_{N, I} =
- A\ n\ x$ is thus given by
\begin{align}
  A = 2 E_c^0 \frac{C_N^0 V_N}{e d}
    = \frac{C_N^0}{C_\Sigma} \frac{e V_N}{d} \;.
\end{align}
and can be tuned by changing the gate voltage $V_N$. Note that we assume
an experimental setup where the charge density on the dot can be
changed by a (plunger) gate $V_G$ and there is an additional gate
voltage from the oscillator $V_N$.

The dimensionless quantity $\ti{A}$ in our notation is thus
\begin{align}
  \ti{A} = \frac{A}{\hbar \Omega} x_0
         = \frac{C_N^0}{C_\Sigma} \frac{e V_N}{\hbar \Omega} \frac{x_0}{d}\;.
\end{align}
With the parameters from Ref.~[\onlinecite{naik06}], for instance, we
can get some very rough estimates for the renormalized quantities of
the SSET. We find that the parameters are of the order of
\begin{align}
  \ti{\Gamma}_\alpha &= 2.5 - 17\;, \\
  \ti{J}_\alpha &= 1 - 2.5\;, \\
  \ti{A} &= 0.001 - 0.01 \;.
\end{align}

\section{Mean-field equations for thermal-oscillator approximation}
\label{app:thosc}

Using the thermal-oscillator approximation in Eq.~\eqref{eq:master}
for the quantities $\expcts{\hat p_{kj}}{}$ we get a closed set of
equations for
\begin{align}
 \ket{\mathbf p} =
   \Big(&
  \expct{\hat p_{-1,-1}}{} \ \expct{\hat p_{1,1}}{} \ \expct{\hat p_{1,-1}}{} \notag \\
& \expct{\hat p_{-1,1}}{} \ \expct{\hat p_{2,2}}{} \ \expct{\hat p_{2,0}}{} \ \expct{\hat p_{0,2}}{} 
 \Big)^T\;,
\end{align}
where $\expcts{\hat p_{00}}{} = 1 - \expcts{\hat p_{-1,-1}}{} - \expcts{\hat p_{1,1}}{} - \expcts{\hat p_{2,2}}{}$ is determined by the normalization of the density, $\Tr[\rho] = 1$. $\ket{\mathbf p}$ fulfills a matrix equation
\begin{align}\label{eq:thosc_p}
  \frac{d}{d\ti{t}} \ket{\mathbf p} &= - \mathbf{M} \ket{\mathbf p} 
                                       + i \ti{J_L} \ket{c}\;,
\end{align}
where $\ket{c} = (0, 0, 0, 0, 0, 1, -1)^T$
and the evolution matrix $\mathbf{M}$ is defined as
\begin{widetext}
\begin{align}
 \mathbf{M} &= 
 \begin{pmatrix}
   \ti{\Gamma_L} & . & - i \ti{J_R} & i \ti{J_R} & . & . & . \\
   . & . & i \ti{J_R} & - i \ti{J_R} & - \ti{\Gamma_R} & . & . \\
   - i \ti{J_R} & i \ti{J_R} & \ti{\Gamma_L}/2 + i \epsilon_{1, -1} & . & . & . & .\\
   i \ti{J_R} & - i \ti{J_R} & . & \ti{\Gamma_L}/2 - i \epsilon_{1, -1} & . & . & .\\
  . & . & . & . & \ti{\Gamma_R} & i \ti{J_L} & - i \ti{J_L} \\
  i \ti{J_L} & i \ti{J_L} & . & . & 2 i \ti{J_L} & \ti{\Gamma_R}/2 + i \epsilon_{20} & . \\
  - i \ti{J_L} & - i \ti{J_L} & . & . & - 2 i \ti{J_L} & . & \ti{\Gamma_R}/2 - i \epsilon_{20}
 \end{pmatrix} \;,
\label{eq:M}
\end{align}
\end{widetext}
where $.$ stands for the entry $0$.
As discussed in the main text we normalize the SSET quantities with
respect to resonator properties, hence the quasiparticle tunneling
$\ti{\Gamma_\alpha} = \Gamma_\alpha/\Omega$, the Cooper pair tunneling
$\ti{J_\alpha} = J_\alpha/2\hbar \Omega$ and the resonance energies
\begin{align}
  \epsilon_{1,-1} &= \frac{4 E_C}{\hbar \Omega} n_0 
                              + \frac{e V}{\hbar \Omega}
                               - 2 \ti{A} \expct{\tx}{}\;, \\  
  \epsilon_{2,0} &= \frac{4 E_C}{\hbar \Omega} (n_0 + 1) 
                               - \frac{e V}{\hbar \Omega}
                               - 2 \ti{A} \expct{\tx}{} \;,
\end{align}
are shifted due to the coupling to the resonator in the thermal
approximation. As can be seen in these expressions, the resonator is
only coupled to the off-diagonal terms in the density matrix $\rho$,
which correspond to Cooper pair tunneling events.

Some assumptions are made to simplify to expressions for
$\epsilon_{2,0}$ and $\epsilon_{1,-1}$.
We assume that the SSET is tuned to the DJQP resonance
with $n_0^{\mathrm{DJQP}} = - 1/2$
and therefore we find deviations from it by
\begin{align}
  \frac{4 E_C}{\hbar \Omega} n_0^0 &= 
    \frac{4 E_C}{\hbar \Omega} n_0^{\mathrm{DJQP}} 
 +  2 \frac{C_G}{C_\Sigma} \Delta \frac{e V_G}{\hbar \Omega} \;,
\end{align}
where $V_G$ is the gate voltage measured from $V_G^{DJQP}$ (symmetric
bias voltage assumed). Furthermore there is a fixed value of the bias
voltage $V^{DJQP}$ which is chosen such that
\begin{align}
2  \frac{eV^{DJQP}}{\hbar \Omega} = \frac{4 E_C}{\hbar \Omega} \;,
\end{align}
and we denote any shift away from $V^{DJQP}$ as $\Delta V$. Consequently the
energy detuning is given by the gate and bias voltage as
\begin{align}
  \epsilon_{1,-1} &= 2 \frac{C_G}{C_\Sigma} \Delta \frac{e V_G}{\hbar \Omega}
  + \Delta \frac{e V}{\hbar \Omega}  - 2 \ti{A} \expct{\tx}{} \;,\\
  \epsilon_{2,0} &= 2 \frac{C_G}{C_\Sigma} \Delta \frac{e V_G}{\hbar \Omega}
  - \Delta \frac{e V}{\hbar \Omega}  - 2 \ti{A} \expct{\tx}{} \;.
\end{align}
For convenience only, it is assumed in the calculations
that $\Delta e V_G/\hbar \Omega$ and
$\Delta e V/\hbar \Omega$ have the same prefactors 
(see Eqs.~\eqref{eq:def_e1m1} and \eqref{eq:def_e20}) and for
abbreviation also $\Delta$ is mostly skipped.

In the stationary limit, i.e. $d/dt \ket{\mathbf p} = 0$, we find immediately
\begin{align}
  \ket{\mathbf p} = i \ti{J_L} \mathbf{M}^{-1} \ket{c}  \;,
\end{align}
or written down explicitly
\begin{align}
\ket{\mathbf p} &= \frac{1}{2}
  \left( \frac{1}{\gamma_R} + \frac{1}{\gamma_L} 
      + \frac{1}{\ti{\Gamma_R}} + \frac{1}{\ti{\Gamma_L}} \right)^{-1} \notag \\
&\times
    \begin{pmatrix}
    \frac{1}{\ti{\Gamma_L}} 
\\[.25cm]
    \frac{1}{\ti{\Gamma_L}} + \frac{2}{\gamma_R}  
\\[.25cm]
    - i \ti{J_R} \frac{2}{\gamma_R} 
         \frac{1}{\frac{\ti{\Gamma_L}}{2} + i \epsilon_{1,-1}} 
\\[.25cm]
      i \ti{J_R} \frac{2}{\gamma_R} 
        \frac{1}{\frac{\ti{\Gamma_L}}{2} - i \epsilon_{1,-1}} 
\\[.25cm]
    \frac{1}{\ti{\Gamma_R}} 
\\[.25cm]
    i \ti{J_L} \frac{2}{\gamma_L}
       \frac{1}{\frac{\ti{\Gamma_R}}{2} + i \epsilon_{2,0}} 
\\[.25cm]
    - i \ti{J_L} \frac{2}{\gamma_L} 
         \frac{1}{\frac{\ti{\Gamma_R}}{2} - i \epsilon_{2,0}}
  \end{pmatrix}\;.  
\end{align}
This result can be inserted in Eqs.~\eqref{eq:current_D} and \eqref{eq:current_ND} and we find the result for the current $\expcts{I}{}$, Eq.~\eqref{eq:current_rate}.

Note that $\expcts{I^{ND}}{} = 2 \expcts{I^{D}}{}$ follows immediately from the equation~\eqref{eq:thosc_p} 
\begin{align}
  \frac{d}{d\ti{t}} \expct{\hat p_{11}}{} &=
 - i \ti{J_R} \left( \expct{\hat p_{1,-1}}{} - \expct{\hat p_{-1,1}}{} \right)
 + \ti{\Gamma_R} \expct{\hat p_{2,2}}{} 
\end{align}
in the stationary limit. This expression is due to the nature of the DJQP process and does not change if we go to more sophisticated approximations.

The shift of the oscillator resonance is proportional to the charge on the island, 
\begin{align}
  \expct{n}{} 
&= 2 \expct{\hat p_{2,2}}{} + \expct{\hat p_{1,1}}{} - \expct{\hat p_{-1,-1}}{} \notag \\
&= \left( \frac{1}{\gamma_R} + \frac{1}{\ti{\Gamma_R}} \right) / 
  \left( \frac{1}{\gamma_R} + \frac{1}{\gamma_L} 
      + \frac{1}{\ti{\Gamma_R}} + \frac{1}{\ti{\Gamma_L}} \right) \notag \\
&=  \left( \frac{1}{\gamma_R} + \frac{1}{\ti{\Gamma_R}} \right)
   \frac{2\expct{I}{}}{3(-e)\Omega}\;.
\end{align}
As obvious from Eq.~\eqref{eq:current_rate} the slowest rate limits
the current and thus determines its value. In the symmetric case
$\ti{J_L} = \ti{J_R} = \ti{J}$ and $\ti{\Gamma_L} = \ti{\Gamma_R} =
\ti{\Gamma}$ the current can also be written more explicitly in terms
of the system parameters \cite{clerk02}
\begin{align}
  \expct{I}{} 
&= (-e) \Omega
   \frac{6 \ti{\Gamma} \ti{J}^2 }
        {2 \epsilon_{2,0}^2 + 2 \epsilon_{1,-1}^2 + 8 \ti{J}^2 + \ti{\Gamma}^2}
\label{eq:current_explicit} \;.
\end{align}
As discussed in the main text the shift due to the coupling to the
leads can be read off from
\begin{align}
\frac{3 (-e) \Omega}{2}
& \Big( \frac{1}{\expct{I}{}} - \frac{1}{\expct{I}{}(\ti{A} = 0)} \Big) 
\label{eq:app_1overcurrent} \\
&= \frac{1}{\gamma_L(\tx)} - \frac{1}{\gamma_L(\tx = 0)}
 + \frac{1}{\gamma_R(\tx)} - \frac{1}{\gamma_R(\tx = 0)}\;.\notag
\end{align}
In the symmetric case $J_L = J_R \equiv J$ and $\Gamma_L = \Gamma_R \equiv \Gamma$ as assumed in the main text, this can be written more explicitly as
\begin{align}
 \frac{3 (-e) \Omega}{2} &
\Big( \frac{1}{\expct{I}{}} - \frac{1}{\expct{I}{}(\ti{A} = 0)} \Big) \notag \\
&= 2 (2 \ti{A} \expct{\ti x}{}) 
  \left( 2 \ti{A} \expct{\ti x}{} - 2 \frac{e V_G}{\hbar \Omega} \right)\;.
\end{align}
Since $\expcts{\tx}{}$ depends on the bias voltage $V$, the value
of $2 \ti{A} \expcts{\tx}{}(V, V_G)$ can be extracted
from the intersection with the voltage axis when plotting $1/\expcts{I}{}(\ti{A})
- 1/\expcts{I}{}(\ti{A} = 0)$.

\section{Mean-field equations for Gaussian approximation}
\label{app:Gaussian}

In the thermal-oscillator approximation we assumed that $\cumu{nx}{}
= \cumu{nv}{} = 0$. For stronger coupling, the accuracy of this
assumption get worse and we proceed to calculate $\cumu{nx}{} = \langle
n \ket{\ket{x \mathbf{p}}}$ with $\bra{n} = (-1, 1, 0, 0, 2, 0, 0)$ in
the Gaussian approximation. Here $\ket{\ket{x \mathbf{p}}}$ denotes the
vector of cumulants $\expct{\expct{x \hat{p}_{kj}}{}}{} = 
\expct{x \hat{p}_{kj}}{} - \expct{x}{} \expct{\hat{p}_{kj}}{}$.
Higher-order expectation values
are assumed to vanish, for example from $\cumu{x^2 \hat p_{kj}}{} = 0$
it follows that
\begin{align}
  \expct{x^2 \hat p_{kj}}{} 
= 2 \expct{x}{} (\expct{x \hat p_{kj}}{} - \expct{x}{} \expct{\hat p_{kj}}{})
  + \expct{x^2}{} \expct{\hat p_{kj}}{}\;.
\end{align}
This approximation leads us to the following set of equations:
\begin{align}
  \frac{d}{d\ti{t}} \ket{\mathbf p} 
&= - \mathbf{M} \ket{\mathbf p} + i \ti{J_L} \ket{c} 
   + 2 i \ti{A} \mathbf{K}_- \ket{\ket{\tx\mathbf p}} \;,\\
  \frac{d}{d\ti{t}} \ket{\ket{\tx \mathbf p}} 
&= - \mathbf{M} \ket{\ket{\tx \mathbf p}} + \ket{\ket{\tv \mathbf p}}
   + 2 i \ti{A} \cumu{\tx^2}{}
         \mathbf{K}_- \ket{\mathbf p} \;,\\
  \frac{d}{d\ti{t}} \ket{\ket{\tv \mathbf p}} 
&= - \mathbf{M} \ket{\ket{\tv \mathbf p}} - \ket{\ket{\tx \mathbf p}}
   - \tgext \ket{\ket{\tv \mathbf p}}
\notag \\ & \quad
   + 2 \ti{A} (\mathbf{K}_+ - \expct{n}{}) \ket{\mathbf p} \;,
\end{align}
where the coupling matrices are given by
\begin{align}
  \mathbf{K}_+ &= \mathrm{diag}\left( -1, 1, 0, 0, 2, 1, 1 \right)
\label{eq:Kplus} \;, \\
  \mathbf{K}_- &= \mathrm{diag}\left( 0, 0, 1, -1, 0, 1, -1 \right)
\label{eq:Kminus} \;.
\end{align}
This set of equations can be solved in the stationary limit 
and is given here in matrix representation
\begin{widetext}
\begin{align}
  | \mathbf{p} \rangle &= i \tilde{J}_L \left[ 
    \mathbf{M}
  + i (2 \tilde{A})^2 \mathbf{K}_- 
    \big( \mathbf{1} + \ti{\gamma}_{ext} \mathbf{M} + \mathbf{M}^2 \big)^{-1}
    \left\{ 
      (\mathbf{K}_+ - \langle n \rangle) 
      + i \langle\langle \tx^2 \rangle\rangle 
        (\ti{\gamma}_{ext} + \mathbf{M}) \mathbf{K}_-
    \right\}
\right]^{-1} | c \rangle \;, \\
  || \tx\mathbf{p} \rangle\rangle &= 2 \tilde{A}
    \big( \mathbf{1} + \ti{\gamma}_{ext} \mathbf{M} + \mathbf{M}^2 \big)^{-1}
   \left[ 
      (\mathbf{K}_+ - \langle n \rangle) 
      + i \langle\langle \tx^2 \rangle\rangle 
        (\ti{\gamma}_{ext} + \mathbf{M}) \mathbf{K}_-
  \right]
  | \mathbf{p} \rangle  \;.
\end{align}
Since in the stationary limit
\begin{align}
  \langle n || \tv \mathbf{p} \rangle\rangle 
&= \langle n | \mathbf{M} || \tx \mathbf{p} \rangle\rangle
\end{align}
we do not need to calculate $\cumu{n\tv}{}$ and find instead
\begin{align}
  \cumu{ \tx^2 }{} &= \cumu{ \tv^2 }{}
+ 2 \tilde{A} \langle n || \tx \mathbf{p} \rangle\rangle 
= 2 \tTB 
+ 2 \tilde{A}/\tgext \langle n | \mathbf{M} || \tx \mathbf{p} \rangle\rangle
+ 2 \tilde{A} \langle n || \tx \mathbf{p} \rangle\rangle \;.
\end{align}
Using the general solution for $\ket{\ket{\tx\mathbf{p}}}$ we can 
rewrite this equation with the expression
\begin{align}
  \langle\langle \tx^2 \rangle\rangle &= 
\frac{2 \tgext \tTB + 2 \gSSET \TSSET(\tx, \tx^2)}
     {\tgext + \gSSET(\tx, \tx^2)} \;,
\end{align}
where we define
\begin{align}
  \gSSET(\tx, \tx^2) &= i (2 \tilde{A})^2 
    \langle n | \big(\mathbf{1} - \tgext (\tgext + \mathbf{M}) \big)
       \big( \mathbf{1} + \tgext \mathbf{M} + \mathbf{M}^2 \big)^{-1}
       \mathbf{K}_- | \mathbf{p}(\tx, \tx^2) \rangle \notag \;, \\
  2 \gSSET \TSSET(\tx, \tx^2) &= (2 \tilde{A})^2 
    \langle n |  \big( \tgext + \mathbf{M} \big) 
       \big( \mathbf{1} + \tgext \mathbf{M} + \mathbf{M}^2 \big)^{-1}
       \big( \mathbf{K}_+ - \langle n \rangle \big) | \mathbf{p}(\tx,
       \tx^2) \rangle  \notag \;.
\end{align}
\end{widetext}
The equation for $\cumu{\tx^2}{}$ has to be solved together with the
expression for $\expcts{\tx}{}$ which leads in some parameter regime
to more than one solution. This bistability of the model is discussed
in detail in the main text.

\section{Charge noise of the SSET}
\label{app:charge}

To calculate the charge noise in the DJQP we evaluate the expression
\begin{align}
  S_n(\omega) &= \int_{-\infty}^\infty dt e^{i \omega t} \cumu{ n(t) n(0) }{}
\\
  &= \int_0^\infty dt \cos(\omega t) 
       \Big( \cumu{ n(t) n(0) }{} + \cumu{ n(- t) n(0) }{} \Big) 
\notag \\ &
   + i \int_0^\infty dt \sin(\omega t) 
       \Big( \cumu{ n(t) n(0) }{} - \cumu{ n(- t) n(0) }{} \Big) \notag
\end{align}
and take into account that $S_n(t)$ is not necessarily
symmetric. Therefore we introduce two new functions \cite{choi03}
\begin{align}
  \chi(t) &= \Tr_{\mathrm{osc}} \Tr_{n_R} \{ e^{- i H t/\hbar} n \rho e^{i H t/\hbar} \} \;, \\
  \eta(t) &= \Tr_{\mathrm{osc}} \Tr_{n_R} \{ e^{- i H t/\hbar} \rho n e^{i H t/\hbar} \} \;.
\end{align}
The partial traces are taken over the oscillator degree of freedom and the tunneled charge $n_R$. These functions have the properties $\cumu{n(t) n(0)}{} = \Tr_n \{ n \chi(t) \}
- \expcts{ n }{}^2$ and $\cumu{n(-t) n(0)}{} = \Tr_n \{ n \eta(t) \} -
\expcts{ n }{}^2 $ where $\Tr_n$ denotes the trace of the charge states of the island. The functions $\chi(t)$ and $\eta(t)$ fulfill the
same differential equation
\begin{align}
  \frac{d}{d\ti{t}} \ket{ \chi } &= - \mathbf{M} | \chi \rangle
                                 + i \tilde{J_L} \expct{n}{} | c \rangle  \;,
\\
 \frac{d}{d\ti{t}} \ket{ \eta } &= - \mathbf{M} | \eta \rangle
                               + i \tilde{J_L} \expct{n}{} | c \rangle \;,
\end{align}
with different initial conditions
\begin{align}
  | \chi(0) \rangle &= \Tr_{\mathrm{osc}} \Tr_{n_R} \{ n \rho \} 
            = \mathbf{K}_\chi | \mathbf{p} \rangle  \;, \\ 
  | \eta(0) \rangle &= \Tr_{\mathrm{osc}} \Tr_{n_R} \{ \rho n \} 
            = \mathbf{K}_\eta | \mathbf{p} \rangle \;,
\end{align}
where $\mathbf{K}_\chi = \mathbf{K}_+ + \mathbf{K}_-$ and $\mathbf{K}_\eta = \mathbf{K}_+ - \mathbf{K}_-$. Straightforwardly we find
\begin{align}
 \cumu{ n(t) n(0) }{} + \cumu{ n(-t) n(0) }{} 
&= 2 \bra{ n } e^{- \mathbf{M} \ti{t}} \left( \mathbf{K}_+ - \expct{n}{} \right)
                 | \mathbf{p} \rangle \notag   \\
 \cumu{ n(t) n(0) }{} - \cumu{ n(-t) n(0) }{} 
&= 2 \bra{ n } e^{- \mathbf{M} \ti{t}} \mathbf{K}_- | \mathbf{p} \rangle  \;,
\end{align}
and consequently
\begin{align}
  S_n(\omega) &= 
  2 \bra{ n }\frac{\mathbf{M}}{(\omega/\Omega)^2 + \mathbf{M}^2} 
                \left( \mathbf{K}_+ - \expct{n}{} \right) | \mathbf{p} \rangle   \notag
\\ &
+ 2 i \bra{ n } \frac{(\omega/\Omega)}{(\omega/\Omega)^2 + \mathbf{M}^2} \mathbf{K}_- | \mathbf{p} \rangle \;.
\end{align}
For example the effective damping and temperature in linear-response theory 
is related to 
\begin{align}
  \frac{S_n(\omega) - S_n(- \omega)}{2 (\omega/\Omega)} &= 
2 i \langle n | \frac{1}{(\omega/\Omega)^2 + \mathbf{M}^2} \mathbf{K}_- | \mathbf{p} \rangle \;, \\
  \frac{S_n(\omega) + S_n(- \omega)}{2} &= 
2 \langle n | \frac{\mathbf{M}}{(\omega/\Omega)^2 + \mathbf{M}^2}
 \left( \mathbf{K}_+ - \expct{n}{} \right) | \mathbf{p} \rangle \;,
\end{align}
as is discussed in more detail in the main text. 
Note the striking similarity to the self-consistency equation for
the Gaussian approximation, Eqs.~(\ref{eq:x2_Gaussian}-\ref{eq:TSSET}).

\section{Calculation of the current noise}
\label{app:noise}

We want to calculate $d/dt \expct{\expct{n_R^2(t)}{}}{}$ and study therefore
the time-evolution of
\begin{align}
       d/dt\ \expct{ n^2_R(t) }{}
&= \Tr\big\{ \dot \rho(t) n_R^2 \big\} \\
&= - \frac{i}{\hbar} \Tr\big\{ ( \rho n_R + 
n_R \rho)\, [ n_R, H] \big\} \notag \;.
\end{align}
Therefore we introduce the 
$n_R$-resolved density matrix. Defining the operator
\begin{align}\label{defpnR}
  \hat p_{kj}^{n_R', n_R} &= \ket{j}\bra{k} \otimes \ket{n_R}\bra{n_R'}
\end{align}
such that $\expcts{\hat p_{kj}}{} = \Tr_{\mathrm{osc}} \Tr_{n_R} \Tr_n \{ \rho | j
\rangle \langle k | \otimes | n_R \rangle \langle n'_R | \}$ and
$\expcts{n_R \hat p_{kj}}{} = \Tr_{\mathrm{osc}} \Tr_{n_R} \Tr_n \{ \rho n_R
| j \rangle \langle k | \otimes | n_R \rangle \langle n'_R | \}$. Note
that $\expcts{\hat p_{kj} n_R}{}$ is a different quantity since $n_R$ and
$\hat p_{kj}^{n_R', n_R}$ do not commute.

We assume the stationary limit in the sense that $\expcts{\hat p_{kj}}{}$
does not depend on time. Then, the $n_R$-resolved density operator can
be calculated in the thermal-oscillator approximation where
the differential equation reads
\begin{align}
  \frac{d}{d\ti{t}}\ \ket{n_R \mathbf{p}}(t) &= - \mathbf{M} \ket{n_R \mathbf{p}} 
+ \ket{A}
+ \expct{ n_R }{} i \tilde J_L \ket{ c } \;,
\\
\ket{A} &= 
\begin{pmatrix}
- 2 i \ti{J_R} \expct{\hat p_{-1, 1}}{} \\
\ti{\Gamma}_R \expct{\hat p_{22}}{} + 2 i \ti{J_R} \expct{\hat p_{1, -1}}{} \\
- 2 i \ti{J_R} \expct{\hat p_{1, 1}}{} \\
2 i \ti{J_R} \expct{\hat p_{-1, -1}}{} \\
0 \\ 0 \\ 0
\end{pmatrix} \;,
\end{align}
since $(-e) \expcts{n_R}{} = \expcts{I}{} t$ is explicitly time-dependent 
(in the stationary limit $\expcts{I}{}(t) \equiv \expcts{I}{}$), 
the differential equation has to be solved explicitly and we find
\begin{align}
  \ket{\ket{n_R \mathbf{p} }}(\ti{t}) &= 
e^{- \mathbf{M} \ti{t}} \ket{\ket{ n_R \mathbf{p}(\ti{t} = 0)}} \\
&
+ (1 - e^{- \mathbf{M} \ti{t}}) \mathbf{M}^{-1} \ket{A}  
- \frac{\expct{I}{}}{(-e)} \mathbf{M}^{-1} \ket{\mathbf{p}}\;.\notag
\end{align}
Note that $\ket{n_R \mathbf{p}}$ denotes similarily to previous
definitions the vector of finite $\expct{n_R \hat p_{kj}}{}$
and $\ket{\ket{n_R \mathbf{p}}} 
= \ket{n_R \mathbf{p}} - \expct{n_R}{} \ket{\mathbf{p}}$.

The noise itself has a structure similar to the current, but with some
correction terms from the counting of $n_R$
\begin{align}
  \expct{I}{} &= (-e) \langle b \ket{\mathbf{p}} \;,\\
  d/d\ti{t} \cumu{n_R^2(\ti{t})}{} &= 2 (-e)^2 \langle b \ket{\ket{n_R \mathbf{p}}} 
+ (- e)^2 \ti{\Gamma}_R \expct{\hat p_{22}}{} \\ &
- 2 (- e)^2 2 i \ti{J}_R \left( \expct{\hat p_{1,-1}}{} + \expct{\hat p_{-1,1}}{} \right)\notag \;, \\
  \bra{b} &=
  \begin{pmatrix}
    0 & 0 & 2 i \ti{J}_R& - 2 i \ti{J}_R & \ti{\Gamma}_R & 0 & 0 
  \end{pmatrix} \;.
\end{align}
If we apply MacDonald's formula \cite{macdonald49} for the symmetrized
noise, i.e.~$S_I(t) = S_I(-t)$, we find [neglecting a factor $e^2$]
\begin{align}
  S_I(\omega) =& S_I(\omega = 0) 
- 2 \bra{b} \mathbf{M}^{-1} \frac{(\omega/\Omega)^2}{(\omega/\Omega)^2 + \mathbf{M}^2} \ket{A}
\notag  \\ &
- 2 \bra{b} \frac{(\omega/\Omega)^2}{(\omega/\Omega)^2 + \mathbf{M}^2} \ket{n_R\mathbf{p}(t=0)} \;,
\\
S_I(\omega = 0) =& 2 \bra{b} \mathbf{M}^{-1} \left( \ket{A}  - \frac{\expct{I}{}}{(-e)} \ket{\mathbf{p}} \right) \\ &
+ \ti{\Gamma}_R \expct{\hat p_{22}}{} 
- 4 i \ti{J}_R \left( \expct{\hat p_{1,-1}}{} + \expct{\hat p_{-1,1}}{} \right)\notag
\;.
\end{align}
The explicit expression for $S_I(\omega = 0)$ is given in the main
text.


\end{document}